\documentclass[10pt,final]{IEEEtran}
\usepackage{graphicx}
\makeatletter
\def\ps@headings{%
\def\@oddhead{\mbox{}\scriptsize\rightmark \hfil \thepage}%
\def\@evenhead{\scriptsize\thepage \hfil \leftmark\mbox{}}%
\def\@oddfoot{}%
\def\@evenfoot{}}
\makeatother
\usepackage[cmex10]{amsmath}
\usepackage{cite,setspace,url}
\usepackage{latexsym}
\usepackage{verbatim}

\usepackage[cmex10]{amsmath}
\usepackage{amssymb}

\newtheorem{lemma}{Lemma}
\newtheorem{definition}{Definition}
\newtheorem{theorem}{Theorem}

\usepackage{mathtools}
\usepackage{texdef2015}
\usepackage[normalem]{ulem}
\usepackage[makeroom]{cancel}
\allowdisplaybreaks 

\newcommand{\NEW}[1]{{\leavevmode\color{blue}{#1}}}
\newcommand{\BLACK}{\color{black}}
\newcommand{\BLUE}{\color{blue}}

\renewcommand{\NEW}[1]{#1}
\renewcommand{\BLACK}{}
\renewcommand{\BLUE}{}

\newlength{\swwidth}
\newcommand{\stw}[2]{\text{\settowidth{\swwidth}{\text{\ensuremath{#1}}}\makebox[\swwidth]{#2}}}

\newcommand{\xz}[1][0]{\stw{x_{0}}{#1}}

\newcommand{\Lcal}{\mathcal{L}}
\newcommand{\Qcal}{\mathcal{Q}}
\newcommand{\Acal}{\mathcal{A}}

\newcommand{\R}{\mathbb{R}}
\newcommand{\dq}[2][q]{\delta_{#2,#1}}
\newcommand{\dqt}[2][q(t)]{\delta_{#2,#1}}
\newcommand{\ql}{q_l}

\newcommand{\qmax}{q_{\max}}
\newcommand{\laml}[1][l]{\lambda^{(#1)}}

\newcommand{\psim}[2][m]{\psi^{(#1)}_{#2}}
\newcommand{\psis}[2][s]{\psi^{(#1)}_{#2}}
\newcommand{\Lpsis}[2][s]{L\psis[#1]{#2}}

\newcommand{\vm}[2][m]{v^{(#1)}_{#2}}
\newcommand{\vs}[2][s]{v^{(#1)}_{#2}}
\newcommand{\vmdot}[2][m]{\dot{v}^{(#1)}_{#2}}
\newcommand{\vsdot}[2][s]{\dot{v}^{(#1)}_{#2}}

\newcommand{\vvm}[2][m]{\vv^{(#1)}_{#2}}
\newcommand{\vvs}[2][s]{\vv^{(#1)}_{#2}}
\newcommand{\vvmdot}[2][m]{\dot{\vv}^{(#1)}_{#2}}
\newcommand{\vvsdot}[2][s]{\dot{\vv}^{(#1)}_{#2}}
\newcommand{\vvmbar}[2][m]{\bar{\vv}^{(#1)}_{#2}}
\newcommand{\vvsbar}[2][s]{\bar{\vv}^{(#1)}_{#2}}

\newcommand{\Lpsim}[2][m]{L\psim[#1]{#2}}

\newcommand{\Lamm}[2][m]{\Lambda^{(#1)}_{#2}}
\newcommand{\Lams}[2][s]{\Lambda^{(#1)}_{#2}}
\newcommand{\cvec}[1]{\Bracket{#1}}

\newcommand{\rowvec}[1]{[\begin{matrix} #1\end{matrix}]}
\newcommand{\rvec}[1]{\rowvec{#1}}

\newcommand{\smeval}[1]{#1|}
\newcommand{\smallset}[1]{\{#1\}}

\newcommand{\mubar}{\bar{\mu}}

\newcommand{\zerov}[1][\mbox{}]{\mathbf{0}_{#1}}
\newcommand{\onev}[1][n]{{\mathbf{1}}_{#1}}

\renewcommand{\vec}[1]{\begin{bmatrix} #1\end{bmatrix}}

\newcommand{\first}{1}
\newcommand{\nlen}{n} 
\newcommand{\range}[2][\first]{#1\,{:}\,#2}
\usepackage{tikz}
\usetikzlibrary{automata,arrows,positioning,calc,shapes.multipart,chains,shapes}

\usepackage{ifthen}

\newcommand{\Others}[1]{\ifthenelse{\equal{#1}{1}}{2}{1}}

\newcounter{lettercount}

\begin{document}

\newboolean{mytwocolumn}
\setboolean{mytwocolumn}{false}

\author{Roy D.~Yates,~\IEEEmembership{Fellow,~IEEE}%
\thanks{Roy Yates is with WINLAB and the ECE Department, Rutgers University, NJ, USA, e-mail: ryates@winlab.rutgers.edu.}
\thanks{This work was presented in part at the 2018 IEEE Infocom Age of Information Workshop. This version will be (more or less) the same as what will appear in the IEEE Transactions on Information Theory.}%
\thanks{This work was supported by NSF award CCF-1717041.}}

\title{The Age of Information in Networks: Moments, Distributions, and Sampling}

\maketitle
%
\begin{abstract}
A source provides status updates to monitors through a network with state defined by a continuous-time finite Markov chain. An age of information (AoI) metric is used to characterize timeliness by the vector of ages tracked by the monitors. Based on a stochastic hybrid systems (SHS) approach,  first order linear differential equations are derived for  the temporal evolution of both the moments and the  moment generating function (MGF)  of the age vector components. It is shown that the existence of a non-negative fixed point for the first moment  is sufficient to guarantee convergence of all higher order moments as well as a region of convergence for the stationary MGF  vector of the age. The stationary MGF vector is then found for the age on a line network of preemptive memoryless servers. From this MGF, it is found that the age at a node is identical in distribution to the  sum of independent exponential service times. This observation is then generalized to linear status sampling networks in which each node receives samples of the update process at each preceding node according to a renewal point process. For each node in the line, the age is shown to be identical in distribution to a sum of independent renewal process age random variables. 
\end{abstract}   
\begin{IEEEkeywords}
Age of information, queueing systems, communication networks, stochastic hybrid systems, status updating, status sampling network
\end{IEEEkeywords}

\section{Introduction}\label{sec:intro}
With the emergence of cyberphysical systems, real-time status updates have become an important and ubiquitous form of communication. These updates include traffic reports, game scores, security system reports from computers, homes, and offices, video feedback from remote-controlled systems, and even location updates from loved ones. 
These examples  share a common description: a source generates time-stamped status update messages that are transmitted through a communication system to one or more monitors. 

The goal of real-time status updating is to ensure that the status of interest is as timely as possible at each monitor. Ideally, a monitor would receive a status update, typically communicated as a data packet, at the very instant it was generated at a source. If this were possible, a source would simply generate status updates as fast as possible. However, system capacity constraints dictate that the delivery of a status message requires a nonzero and typically random time in the system. 

From these observations, timeliness of status updates has emerged as a new field of network research.  It has been shown \NEW{in simple queueing systems} that timely updating is not the same as maximizing the utilization of the  system that delivers these updates, nor the same as ensuring that updates are received with minimum delay \cite{Kaul-YG-infocom2012}. 
While utilization is maximized by sending updates as fast as possible, this strategy will lead to the monitor receiving delayed updates that were backlogged in the communication system. 
In this case, the timeliness of status updates at the receiver can be improved by {\em reducing} the update rate. 
On the other hand,  throttling the update rate will also lead to the monitor having
unnecessarily outdated status information because of a lack of updates.

This has led to the introduction of new performance  metrics, based on the {\em Age of Information (AoI)}, that describe the timeliness of one's knowledge of an entity or process. Specifically, an update packet with time-stamp $u$ is said to have age $t-u$ at a time $t\ge u$.  When the monitor's freshest\footnote{One update is fresher than another if its age is less.} received update at time $t$ has time-stamp $u(t)$,  the  age is the random process $x(t)=t-u(t)$. Optimization based on AoI metrics of both the network and the senders' updating policies  has yielded new and even surprising results \cite{Sun-UBYKS-IT2017UpdateorWait,Yates-isit2015}.

Nevertheless, the analysis of the time-average AoI has proven challenging, even when simple queues have been used to model update delivery processes. The primary objective of this work is to develop new tools for AoI analysis in networks.  Building on prior work \cite{Yates-Kaul-IT2019} that introduced the stochastic hybrid system (SHS) for AoI analysis, this paper employs SHS to analyze the temporal convergence of higher order AoI moments  and the moment generating function (MGF)  of an AoI process. The MGF enables characterization of the stationary distribution of the AoI in a class of  {\em status sampling} networks, a networking paradigm in which samples of a node's status update process are delivered as a point process to neighbor nodes.  This type of model may be useful in a high speed network in which updates represent small amounts of protocol information (requiring negligible time for transmission)  that are given priority over data traffic.  While the transmission of a single update may be negligible, the update rates are limited so that  protocol information in the aggregate does not consume an excessive fraction of network resources. 

\subsection{AoI Background}
\label{sec:background}

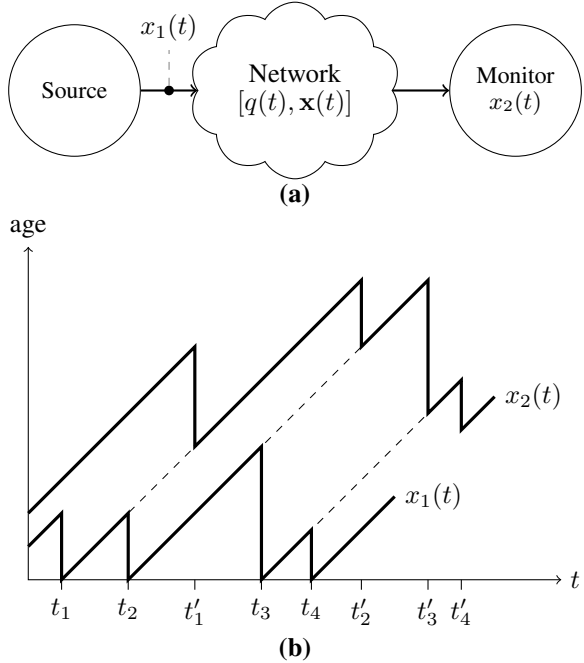
\begin{figure}[t]
\centering
\begin{tabular}[t]{ccc}
\begin{tikzpicture}[baseline=(current bounding box.center),node distance=0.75cm,
dot/.style={draw,circle,fill=black,minimum size=1mm,inner sep=0pt},
circnode/.style={draw,circle,align=center,minimum size=1.75cm}]
\node [circnode] (newsource) {\small Source};
\node  [cloud, draw,cloud puffs=10,cloud puff arc=120, aspect=1.3, inner ysep=1em] [right = of newsource](net){\shortstack{Network\\$[q(t),\xv(t)]$}};
\draw[->,thick] (newsource.east) -- node[dot, pos=0.5, pin={[pin edge={dashed,-}]90:$x_1(t)$}](x1node){}(net.west);
\node [circnode] (localcache)[right = of net] {\small\shortstack{Monitor\\$x_2(t)$}};
\draw[->,thick] (net.east) -- (localcache.west);
\end{tikzpicture}\\[5mm]
{\bf (a)}\\
\begin{tikzpicture}[baseline=(current bounding box.center),scale=\linewidth/20cm]
\draw [<->] (0,10) node [above] {age} -- (0,0) -- (16,0) node [right] {$t$};
\draw [very thick] (0,2) -- (5,7) -- (5,4)  -- (10,9) 
-- (10,7)  -- (12,9) -- (12,5) -- (13,6) -- (13,4.5) -- (14,5.5) node [right] {$x_2(t)$};
\draw [ultra thin, dashed] (1,0) to (10,9) to (10,7) to (3,0);
\draw 
(1,0) -- (1,-0.3) node [below] {$t_1$} 
(3,0) --(3,-0.3) node [below] {$t_2$} 
(5,0) -- (5,-0.3) node [below] {$t'_1$}
(7,0) --(7,-0.3) node [below] {$t_3$}
(8.5,0)--(8.5,-0.3)  node [below] {$t_4$}
(10,0) --(10,-0.3) node [below] {$t'_2$}
(12,0) -- (12,-0.3) node [below] {$t'_3$}
(13,0) -- (13,-0.3) node [below] {$t'_4$};
\draw [very thick] (0,1) -- (1,2) -- (1,0)  -- (3,2) 
-- (3,0)  -- (7,4) -- (7,0) -- (8.5,1.5)--(8.5,0)--(11,2.5) node [right] {$x_1(t)$};
\draw [thin, dashed] (7,0) to (12,5);
\end{tikzpicture}\\
{\bf (b)} 
\end{tabular}
\caption{(a) Fresh updates from a source pass through the network to  a destination monitor.  Monitor $1$ (marked by $\bullet$) sees fresh update  packets at the network access link. (b) Since Monitor $1$ sees fresh updates as a point process at times $t_i$, its age process $x_1(t)$ is reset to zero at times $t_i$.   Since the destination monitor sees updates that are delivered at times $t'_i$ after traveling through the network, its age process $x_2(t)$ is reset to $x_2(t'_i)=t'_i-t_i$, which is the age of update $i$ when it is delivered.}
\label{fig:cloudnet}\end{figure}

An update is said to be {\em fresh} when its timestamp is the current time $t$ and its age is zero. \NEW{As depicted in  Figure~\ref{fig:cloudnet}(a),  the canonical updating model has a source that submits fresh updates to a network that delivers those updates to a destination monitor. In this work, there are additional monitors/observers in the network that serve to track the ages of updates in the network. 

In the Figure~\ref{fig:cloudnet}(a) example, an additional monitor observes the fresh updates as they enter the network.  These fresh updates are submitted at times $t_1,t_2,\ldots$ and this induces the AoI process $x_1(t)$ shown in Figure~\ref{fig:cloudnet}(b). Specifically, $x_1(t)$ is the age of the most recent update seen by a monitor at the input to the network. Because the updates are fresh, $x_1(t)$ is reset to zero at each $t_i$. However, in the absence of a new update, the age $x_1(t)$ grows at unit rate.  
If the source  in Fig.~\ref{fig:cloudnet} submits fresh updates as a renewal point process,  the AoI $x_1(t)$ is simply the age (also known as the backwards excess) \cite{Ross1996stochastic,Gallager2013stochastic} of the renewal process.}

\NEW{These updates pass through a network  and are delivered to the destination monitor at corresponding times $t'_1,t'_2,\ldots$. Consequently, the AoI process $x_2(t)$ at the destination monitor is reset at time $t'_i$ to $x_2(t'_i)=t'_i-t_i$, which is the age of the $i$th update when it is delivered. Once again, absent the delivery of a newer update, $x_2(t)$ grows at unit rate.  Hence the age processes $x_1(t)$ and $x_2(t)$ have the characteristic sawtooth patterns shown in Figure~\ref{fig:cloudnet}(b). Furthermore, any other monitor in the network that sees updates arrive some time after they are fresh, will have a sawtooth age process $x(t)$ resembling that of $x_2(t)$.}

Initial work on age has focused on applying graphical methods to sawtooth age waveforms $x(t)$  to evaluate the limiting time-average AoI
\begin{equation}\eqnlabel{time-average-age}
\Xbar=\limty{T} \frac{1}{T}\int_0^T x(t)\,dt.
\end{equation}
While the time average $\Xbar$ is often referred as the AoI, this work employs AoI and age as synonyms that refer to the process $x(t)$ and call $\Xbar$ the average AoI or average age.\footnote{In Section~\ref{sec:sampling}, it will be necessary to distinguish between the AoI and the age of a renewal process used for random sampling.} 
 
\subsection{Prior Work}
\label{sec:related}
AoI analysis of updating systems started with the analyses of status age in single-source single-server first-come first-served (FCFS) queues \cite{Kaul-YG-infocom2012}, the M/M/1 last-come first-served (LCFS) queue with preemption in service \cite{Kaul-YG-ciss2012}, and the M/M/1 FCFS system with multiple sources \cite{Yates-Kaul-isit2012}.  
Since these initial efforts, there have been a large number of  contributions to AoI analysis. 
 
 To evaluate AoI for a single source sending updates through a network cloud \cite{Kam-KE-isit2013random} or through an M/M/$m$ server \cite{Kam-KE-isit2014diversity,Kam-KNE-IT2016diversity,Yates-isit2018}, out-of-order packet delivery was the key analytical challenge. A related (and generally more tractable) metric, peak age of information (PAoI), was introduced in \cite{Costa-CE-isit2014}. 
Properties of PAoI were also studied for various M/M/1 queues that support preemption of updates in service or discarding of updates that find the server busy \cite{Costa-CE-IT2016management,Kavitha-AS-arxiv2018} or have packet erasures at the queue output \cite{Chen-Huang-isit2016}. In  \cite{Costa-CE-isit2014,Costa-CE-IT2016management}, the authors analyzed AoI and PAoI for  queues that discard arriving updates if the system is full and also for  a third queue in which an arriving update would preempt a waiting update.  

For a single updating source, distributional properties of the age process were analyzed for the D/G/1 queue under FCFS \cite{Champati-AG-aoi2018}, as well as for single server FCFS and LCFS queues  \cite{Inoue-MTT-IT2019}.  Packet deadlines were   found to improve AoI \cite{Kam-KNWE-isit2016deadline,Kam-KNWE-IT2018deadline} and age-optimal preemption policies were identified for updates with deterministic service times \cite{Wang-FY-spawc2018}.

There have also been efforts to evaluate and optimize age for multiple sources sharing a queue or simple network \cite{Yates-Kaul-IT2019,Huang-Modiano-isit2015,Pappas-GKK-icc2015,Kadota-UBSM-Allerton2016,Kaul-Yates-isit2017, Najm-Telatar-aoi2018,Beytur-UB-SIU2018,Jiang-KZZN-isit2018,Jiang-KZZN-iot2019,Kostas-PEA-jcn2019}. \NEW{In \cite{Yates-Kaul-IT2019}, the SHS approach was introduced to extend AoI results to preemptive queues with multiple sources.}  SHS was also used to evaluate age in a CSMA system\cite{Maatouk-AE-ToN2020} as well as NOMA and OMA \cite{Maatouk-AE-aoi2019} systems. With synchronized arrivals, a maximum age first policy was shown to be optimal under preemption in service and near-optimal for non-preemptive service \cite{Sun-UBK-aoi2018}. A similar maximum age matching approach  was analyzed for an orthogonal channel system \cite{Tripathi-Moharir-globecom2017}. Scheduling based on the Whittle index was also shown to perform nearly optimally \cite{Kadota-UBSM-Allerton2016,Jiang-KZN-itc2018,Hsu-isit2018}.  AoI analysis of preemptive priority service systems has also been explored \cite{Najm-NT-IT2020,Kaul-Yates-isit2018priority,Maatouk-AE-isit2019}.
Updates through  communication channels have also been studied, including  fading channels \cite{Huang-Qian-globecom2017,Tang-WSS-allerton2019}, hybrid ARQ  for channels with bit erasures \cite{Parag-TC-wcnc2017realtime,Najm-YS-isit2017,Yates-NSZ-isit2017,Ceran-GG-wcnc2018,Sac-BUBD-spawc2018},  and updates by  energy harvesting sources \cite{Bacinoglu-CUB-ita2015,Bacinoglu-SUBM-isit2018,Wu-YW-green2018,Farazi-KB-aoi2018,Feng-Yang-isit2018,Baknina-Ulukus-arxiv2018coded,Arafa-Ulukus-TW2019,Arafa-YUP-IT2020}.

%
The first evaluation of the average AoI over multihop network routes \cite{Talak-KM-allerton2017} employed a discrete-time version of the  status sampling network described here in Section~\ref{sec:sampling}. 
When multiple sources employ wireless networks subject to interference constraints, AoI has been analyzed under a variety of link scheduling strategies \cite{He-YE-IT2018,Lu-JL-mobicom2018,Talak-KM-2018distributed,Talak-KM-wiopt2018perfectCSI,Talak-KKM-isit2018,Maatouk-AE-ITW2018,Yang-AQP-globecom2019,Buyukates-SU-JCN2019,Leng-Yener-2019TCCN}.  Age bounds were developed from graph connectivity properties  \cite{Farazi-KB-JCN2019} when each node needs to update every other node. For DSRC-based vehicular networks, update piggybacking strategies were developed and evaluated \cite{Kaul-YG-globecom2011piggybacking}. 

When update transmission times over network links are exponentially distributed,  sample path arguments were used \cite{Bedewy-SS-isit2016,Bedewy-SS-isit2017,Bedewy-SS-ToN2019} to show that a preemptive Last-Generated, First-Served (LGFS) policy results in smaller age processes at all nodes of the network than any other causal policy. Note that \cite{Bedewy-SS-ToN2019} and this work can be viewed as complementary in that \cite{Bedewy-SS-ToN2019} proves the  age-optimality of LGFS policies and this work provides analytic tools for the evaluation of those policies.   

In addition to these queue/network analyses,  AoI has also appeared in various application areas,
including  timely updates via replicated servers \cite{Zhong-YS-allerton2017,Sang-LJ-globecom2017,Zhong-YS-aoi2018,Zhong-YS-spawc2018}, timely source coding \cite{Zhong-Yates-dcc2016lossless,Zhong-YS-isit2017,Mayekar-PT-isit2018lossless,Mayeker-PT-arxiv2018}, dissemination of channel state information \cite{Costa-VE-icc2015,Klein-FHB-TW2017,Farazi-KB-icccn2017,Farazi-KB-icassp2016}, differential encoding of temporally correlated updates \cite{Bhambay-PP-wcnc2017}, correlated updates from multiple cameras \cite{HeFD-aoi2018}, UAV trajectory optimization \cite{Abd-Elmagid-Dhillon-vt2019}, periodic updates from correlated IoT sources \cite{Hribar-CKD-globecom2017}, mobile cloud gaming \cite{Yates-THR-infocom2017}, google scholar updating \cite{bastopcu-Ulukus-arxiv2020google}, 
and game-theoretic approaches to network resource allocation for updating sources  \cite{Nguyen-KKWE-wiopt2017,Xiao-Sun-aoi2018,Gopal-Kaul-aoi2018,Garnaev-ZZY-aoi2019}.


\subsection{Paper Overview}\label{sec:overview}

This work is based on the system depicted in Figure~\ref{fig:cloudnet}(a) in which a source sends update packets through a network  to a destination monitor.  \NEW{This system may be a simple queue or it may be a complex network in which updates follow multihop routes to the monitor and the network carries other network traffic  that interferes with the updating process of interest.  This work focuses on a class of systems in which the movements of updates in the network are described by a finite-state continuous-time Markov chain $q(t)$. 
To deal with continuously-growing age processes under a finite number of states,  $q(t)$ is embedded in a stochastic hybrid system (SHS) \cite{Hespanha-2006modelling} with hybrid state $[q(t),\xv(t)]$ such that $\xv(t)$,  the age vector or AoI process, 
 is a real-valued non-negative vector that describes the continuous-time evolution of a collection of AoI processes.}   
 
The SHS approach to age with hybrid state $[q(t),\xv(t)]$ was introduced  in \cite{Yates-Kaul-IT2019} 
where it was shown that age tracking can be implemented as a simplified SHS with non-negative linear reset maps in which the continuous state is a piecewise linear process 
\cite{Vermes-1980,Gnedenko-Kovalenko-1966}, a special case of piecewise deterministic processes \cite{Davis-1984,Deville-DDZ-siam2016moment}. In \cite{Yates-Kaul-IT2019}, the SHS approach led to a system of  first order ordinary differential equations describing the temporal evolution of the expected value of the age process.  
This led to a set of age balance equations and simple conditions \cite[Theorem~4]{Yates-Kaul-IT2019} under which the average age $\E{\xv(t)}$ converges to a fixed point.  
\BLUE
In relation to \cite{Yates-Kaul-IT2019}, this work makes three contributions:
\begin{itemize}
\item The SHS method introduced in \cite{Yates-Kaul-IT2019} for the average age  is generalized to the analysis of higher order moments  and (through the MGF) distributional properties of age processes. 
\item This extended SHS analysis of the MGF is employed to characterize distributional properties of age processes in a class of queueing networks with preemptive servers and memoryless service times.
\item The observations derived for the network of preemptive servers are generalized, in the sense that memoryless service times are supplanted by service times with general distributions. These networks are shown to lend themselves a new description as ``status sampling networks.''
\end{itemize}

Specifically, Section~\ref{sec:newSHSintro} presents Lemma~\ref{lem:pi-vv-derivs},  a system of first order linear differential equations  for the temporal evolution of the higher-order age moments and a moment generating function of the age vector.  The differential equations of Lemma~\ref{lem:pi-vv-derivs} are the foundation for results in Sections~\ref{sec:stationary}, \ref{sec:line-network}, and~\ref{sec:sampling}.
\BLACK
Section~\ref{sec:stationary}  shows how fixed points of these differential equations describe  stationary higher order moments 
and the stationary  MGF of the age process.  In particular, it is shown that a non-negative fixed point for the first moment guarantees the existence of all moments of the age and a region of convergence for the MGF.\footnote{While the higher order moments can be derived from the MGF, there are cases in which direct calculation of the moments is more straightforward.}

These results are first summarized in \Thmref{age-moments} in a form convenient for hand calculation. \NEW{The method of \Thmref{age-moments}  is demonstrated in Section~\ref{sec:MM11abandon} with the analysis of the age MGF of an M/M/1/1 queue that can abandon updates in service.} However, the proof of \Thmref{age-moments} requires a matrix representation of the system of differential equations and its fixed point. Specifically, proof of \Thmref{age-moments-matrix}, the matrix version of \Thmref{age-moments}(a) for the age moments appears in Section~\ref{sec:matrix-moments}   and proof of \Thmref{vectorMGF},  the matrix version of \Thmref{age-moments}(b) for the age MGF, is in Section~\ref{sec:matrix-MGF}.

Section~\ref{sec:line-network} goes on to use  Theorems~\thmref{age-moments-matrix} and~\thmref{vectorMGF} to derive the moments, MGF, and stationary distribution of the age in a line network with memoryless preemptive servers. From the MGF, it is found that the age at a node has stationary distribution identical to a sum of independent exponential random variables. This generalizes  a preliminary result  for the average age that was based on SHS analysis and the method of fake updates \cite{Yates-aoi2018}. 

In Section~\ref{sec:sampling}, it is shown that the structural simplicity of the AoI in the line network derives from the observation that 
preemptive memoryless servers employing fake updates can be viewed as a {\em status sampling network} in which samples of the update process at a node $i$ are delivered to a neighboring node $j$ as a point process.  When this point process is a renewal process,   the AoI at node $j$ is found to have a stationary distribution given by the independent sum of the stationary AoI at node $i$ and the stationary age of the renewal process.

Note that Sections~\ref{sec:stationary}, \ref{sec:line-network}, and~\ref{sec:sampling} follow from the differential equations of Lemma~\ref{lem:pi-vv-derivs}. While Lemma~\ref{lem:pi-vv-derivs} is derived from an SHS model of age processes in a network, employing the system of differential equations in Lemma~\ref{lem:pi-vv-derivs} does not require the reader to tackle the somewhat onerous SHS notation and terminology. Hence the derivation of Lemma~\ref{lem:pi-vv-derivs} is deferred to Section~\ref{sec:SHS} where sawtooth age processes are modeled as a stochastic hybrid system. From fundamental SHS properties, Dynkin's formula is used to derive Lemma~\ref{lem:pi-vv-derivs}, the basis for all subsequent results. The paper concludes in Section~\ref{sec:conclusions}.

\subsection{Notation}
For integers $m\le n$, $\range[m]{n}=\set{m,m+1,\ldots,n}$; otherwise $\range[m]{n}$ is an empty set. The Kronecker delta function $\delta_{i,j}$ equals $1$ if $i=j$ and otherwise equals $0$. 
The vectors $\zerov[n]$ and $\onev[n]$ denote the row vectors $\rowvec{0  &\cdots& 0}$ and  $[\begin{matrix}1 &\cdots &1\end{matrix}]$ in $\R^n$. The $n\times n$ identity matrix is $\Imat_n$.  With $n$ unspecified, $\zerov[]$, $\onev[]$, and $\Imat$ will have dimensions that can be inferred from context. A vector $\xv\in\R^n$ is  a $1\times n$ row vector.
For vector $\xv=\rowvec{x_1,&\cdots&x_n}$, $[\xv]_j=x_j$ denotes the $j$th element. The vector $\ev_i$ denotes the $i$th Cartesian unit vector satisfying $[\ev_i]_j=\delta_{i,j}$.   A matrix $\Bmat$ has $i,j$th element $[\Bmat]_{i,j}$ and $j$th column $[\Bmat]_j$.  In some instances, it will be convenient for vector/matrix indexing to start at $j=0$ such that $\xv=\rowvec{x_0&\cdots&x_{n}}$ and $\Bmat$ has upper left element $[\Bmat]_{00}$ and leftmost column $[\Bmat]_{0}$.  For a process $\xv(t)$,  $\dot{\xv}$ and $\dot{\xv}(t)$  both denote the derivative $d\xv(t)/dt$. For a scalar function $g(\cdot)$ and vector $\xv\in\R^n$, $g(\xv)=\rowvec{g(x_1)& \cdots & g(x_{n})}$. In particular, the vectors $\xv^m=\rowvec{x_1^m&x_2^m&\cdots&x_{n}^m}$ and $e^{s\xv}=\rowvec{e^{sx_1}&\cdots &e^{sx_{n}}}$ appear frequently.

A random variable $X$ has CDF $\icdf{X}=\prob{X\le x}$ and 
a pair of random variables $X,Y$ has joint CDF $\icdf{X,Y}=\prob{X\le x,Y\le y}$. Random variables $X$ and $Y$ having the same distribution is denoted $X\sim Y$. In addition,  $X\oplus Y$ denotes the sum $X+Y$ under the product distribution $\icdf{X,Y}=\icdf{X}\icdf{Y}$.  When $\xv$ is a random vector,  $\E{e^{s\xv}}$ is referred to as the MGF of $\xv$ or the MGF vector.\footnote{This is a restricted form of MGF that is insufficient for describing dependency among the vector components, but will enable us to derive the marginal distribution of each $x_i$ through  $\E{e^{s x_i}}$.} When $x(t)$ is a random process such that the random variable $x(t)$ converges to a stationary distribution, $X(t)$  denotes the random process initialized with this stationary distribution.

\BLUE
\section{A Model for Updates in Networks}\label{sec:newSHSintro}
This section describes a method for evaluating the moments and MGF of an AoI process
$\xv(t)=\rowvec{x_{\first}(t) & \cdots &x_n(t)}$
in updating systems with discrete state  given by a finite-state continuous-time Markov chain $q(t)\in\Qcal=\set{0,\ldots,\qmax}$. 
For example, in the Section~\ref{sec:MM11abandon} age analysis of updates through an M/M/1/1 queue with abandonment, $q(t)\in\set{0,1}$ will be the queue size and $\xv(t)=\rowvec{x_{1}(t)  &x_2(t)}$ will describe the age state at the queue and at the monitor. In the more general example of Figure~\ref{fig:cloudnet}(a), fresh updates are submitted  by a source and pass through a network to a monitor; the movements of update packets through the network are described by $q(t)$. 

In a graphical representation of the Markov chain $q(t)$, each state $q\in\Qcal$ is a node and nonzero transition probabilities are enumerated by an index set $\Lcal$. Each  $l\in\Lcal$  corresponds to  a directed transition $(q_l,q'_l)$  from state $q_l$ to $q'_l$ with fixed transition rate $\laml$ while $q(t) = q_l$. Returning to the M/M/1/1 example, Figure~\ref{fig:MM11abandon} depicts the Markov chain $q(t)$ with states $\Qcal=\set{0,1}$ and transition set $\Lcal=\set{1,2,3}$; transition $l=2$, corresponding  to the directed edge $(q_2,q'_2)=(1,0)$, is a rate $\laml[2]=\mu$ transition from state $q_2=1$ to state $q'_2=0$. 

The ages of the update packets, or equivalently the age processes at various monitors in the network, are described by the continuous state $\xv(t)$.\footnote{For a given system, the specification of the continuous state $\xv(t)$ is not unique. In \cite{Yates-Kaul-IT2019}, these components were explicitly associated with ages of update packets in a system.
In general, when $\xv(t)\in\R^{\nlen}$, one component of $\xv(t)$  is needed to track the age at the monitor and the other  components enable tracking of the ages of up to $\nlen-1$ update packets in the system.}
In this work,  $x_j(t)$ is the age process of a monitor that sees update packets that pass through  a ``position'' or ``observation post'' $j$ in the system. An observation post $j$  may refer to a node $j$ or link $j$ in a network,  a position $j$ in a queue,   or a
server $j$  in a multiple server system. In any case, $x_j(t)$ is the age process at monitor $j$, or simply the age of monitor $j$. In the Figure~\ref{fig:cloudnet}(a) example, $x_1(t)$ is the AoI of a monitor observing fresh updates on the network access link of the source while $x_2(t)$ is the AoI at the destination monitor node. 

Each $x_j(t)$ grows at unit rate in the absence of a more recent update passing through  the observation post. 
Moreover, $x_j(t)$ can make a discontinuous jump only in a transition of $q(t)$. Specifically, in the example of Figure~\ref{fig:cloudnet}, $x_1(t)$ is reset  to zero at time $t_i$ when a fresh update arrives at the network and $x_2(t)$ is reset to $x_2(t'_i)=t'_i-t_i$ at time $t'_i$ when the $i$th update is delivered to the monitor. In general, $x_j(t)$ has a downward jump when a discrete state transition causes an update to pass observation post $j$.

The coupled  evolution of the discrete state $q(t)$ and the continuous state $\xv(t)$ is described by a particularly simple instance of a stochastic hybrid system with the following rules:
\begin{itemize}
\item  As long as the discrete state $q(t)$ is unchanged,
 \begin{equation}\eqnlabel{SHSde}
 \dot{\xv}(t)=\onev[].
 \end{equation}
That is, in each discrete state, the age at each monitor grows at unit rate.  \item When the discrete state $q(t)$ undergoes a transition, a component $x_j(t)$ of the continuous state $\xv(t)$ will make a downward jump if a fresher update is observed by monitor $j$. 
\end{itemize}
These two rules yield sawtooth age processes at each monitor.

The next step is to  describe a model for the downward jumps in the age processes.  For each discrete-state transition $l\in\Lcal$, the downward jumps in the  the continuous state $\xv$  are described by a {\em transition reset map} which is a linear mapping of the form $\xv'=\xv\Amat_l$.
\BLACK  
That is, transition $l$ causes the system to jump from discrete state  $q_l$ to $q'_l$ and resets the continuous state from $\xv$ to $\xv'=\xv\Amat_l$.



For tracking of  AoI processes, each $\Amat_l$ is a binary matrix that has  no more than a single $1$ in a column. Such an $\Amat_l$ will be called an {\em age assignment} matrix.  The set of  age assignments $\set{\Amat_l}$ will depend on the specific queue discipline at each node, the network routing function, and the indexing scheme for updates in the system.
 
Column $j$ of $\Amat_l$ determines how $x'_{j}$ is set when transition $l$ takes place. In particular, if $[\Amat_l]_{i,j}=1$, then transition $l$  resets the  age $x_j$ at  monitor $j$  to  $x'_j=x_i$. This corresponds to monitor $j$ observing the current  update of monitor $i$.  For example, in an FCFS queue in which each queue position has a monitor, this occurs when an update  changes its queue position from $i$ to $j=i-1$ with the service completion of the head-of-line update.  

Another important case is a transition $l$ in which a monitor $j$ observes a fresh update. In this transition,  $x'_{j}=0$ because the update is fresh. This requires $[\Amat_l]_j$, the $j$th column of $\Amat$, to be an all-zero column. In all cases, the  age assignment  $\Amat_l$ encodes  age reductions at those monitors that observe the deliveries of fresher updates. 

To summarize, this work considers networks in which the age process evolves as $\dot{\xv}(t)=\onev[]$ in each discrete state $q\in\Qcal$ and the transitions $l$ are described by  the tuples $a_l=(q_l,q'_l,\laml,\Amat_l)$.  Denoting the  set of transitions as $\Acal=\set{a_l: l\in\Lcal}$,  the tuple  $(\Qcal,\onev[],\Acal)$ defines these AoI processes.
\begin{definition}\label{def:SHS-AoI}
An age-of-information (AoI) SHS $(\Qcal,\onev[],\Acal)$ is an SHS in which the discrete state $q(t)\in\Qcal$ is a continuous-time Markov chain with transitions $l\in\Lcal$ from state $q_l$ to $q'_l$ at rate $\laml$ and the continuous state evolves according to $\dot{\xv}(t)=\onev[]$ in each discrete state $q\in\Qcal$ and is subject to the age assignment map $\xv'=\xv\Amat_l$ in transition $l$. 
\end{definition}

\BLUE
The analysis of the moments and MGF of the age process in an AoI SHS requires an analytical  model that enables tracking of the evolution of the $m$th moment $\E{x_j^m(t)}$ and moment generating function $\E{e^{sx_j(t)}}$ for all age components $x_j(t)$. 
This is done with  a bookkeeping system for the continuous state that employs  the discrete state for partitioning. Recalling that $\delta_{i,j}$ is the Kronecker delta function,  this bookkeeping system defines 
\begin{subequations}\eqnlabel{vqi-defn}
\begin{align}
\vm{\qbar, j}(t)&
=\E{x_j^m(t)\dqt{\qbar}},\eqnlabel{vqi0}\\
\vs{\qbar,j}(t)&
=\Ebig{e^{sx_j(t)}\dqt{\qbar}}
\end{align}
\end{subequations}
for all states $\qbar\in\Qcal$ and age processes $x_j(t)$. These expected value processes are then gathered  in 
\BLACK
the vector functions
\begin{subequations}\eqnlabel{vv-defn}
\begin{align}
 \vvm{\qbar}(t)&=\rvec{\vm{\qbar, \first}(t)&\cdots&\vm{\qbar, n}(t)}=\Ebig{\xv^m(t)\dqt{\qbar}},
 \eqnlabel{vvm-defn}\\
 \vvs{\qbar}(t)&=\rvec{\vs{\qbar,, \first}(t)&\cdots&\vm{\qbar, n}(t)}=\Ebig{e^{s\xv(t)}\dqt{\qbar}}.
 \eqnlabel{vvs-defn}
\end{align}
\end{subequations}
Note that 
$\vvm{\qbar}(t)$ and $\vvs{\qbar}(t)$ use the dummy parameter names $m$ and $s$ to distinguish between classes of test functions. While this is generally undesirable, it will highlight the parallelism in formulations. This abuse of notation can create ambiguity; $\vm[2]{\qbar, j}$ may refer to 
$\smeval{\vm{\qbar, j}}_{m=2}$ or to $\smeval{\vs{\qbar, j}}_{s=2}$.   Hence the convention that  $\vm[i]{\qbar, j}(t)$ for any integer $i\ge1$ refers to 
$\vm{\qbar, j}(t)$ at $m=i$ is maintained.

Conveniently, at $m=0$ or $s=0$, this abuse of notation is consistent in yielding the discrete-state indicator
\begin{align}\eqnlabel{psi0}
\eval{\vm{\qbar, j}(t)}_{m=0}=\eval{\vs{\qbar, j}(t)}_ {s=0}= \E{\dqt{\qbar}}=\prob{q(t)=\qbar}.
\end{align}
Since the index $j$ in \eqnref{psi0} has become redundant, 
\eqnref{vv-defn} implies
\begin{IEEEeqnarray}{rCl}\eqnlabel{vq0-defn}
\vvm[0]{\qbar}(t)=\eval{\vvm{\qbar}(t)}_{m=0}&=&\eval{\vvs{\qbar}(t)}_{s=0}= \onev[n]
\prob{q(t)=\qbar}\IEEEeqnarraynumspace
\end{IEEEeqnarray} 
is a vector of identical copies of $\prob{q(t)=\qbar}$. This vectorized redundancy will 
simplify some subsequent matrix algebra. 

Since $\dqt{\qbar}=0$ for $q(t)\neq \qbar$, it follows from \eqnref{vv-defn} and the law of total expectation that for all $\qbar\in\Qcal$,
\begin{subequations}\eqnlabel{vv-interp}
\begin{align}
 \vvm{\qbar}(t)&=\E{\xv^m(t)|q(t)=\qbar}\prob{q(t)=\qbar},\\ 
 \vvs{\qbar}(t)&=\Ebig{e^{s\xv(t)}|q(t)=\qbar}\prob{q(t)=\qbar}, 
\end{align}
\end{subequations}
\NEW{In \eqnref{vv-interp}, $\vvm{\qbar}(t)$ and $\vvs{\qbar}(t)$ can be interpreted as conditional expectations  of $\xv^m(t)$ and $e^{s\xv(t)}$   given $q(t)=\qbar$, weighted by the state probability $\prob{q(t)=\qbar}$.} Furthermore, since
\begin{align}\eqnlabel{x-decomposition}
\xv^m(t)=\sum_{\qbar\in\Qcal} \xv^m(t) \delta_{\qbar,q(t)},
\end{align}
summing the weighted conditional moments $\vvm{\qbar}(t)$ yields the vector of $m$th moments:
\begin{subequations}
\begin{align}
\E{\xv^m(t)}=\sum_{\qbar\in\Qcal} \E{\xv^m(t)\delta_{\qbar,q(t)}}
=\sum_{\qbar\in\Qcal} \vvm{\qbar}(t).\eqnlabel{mth-moment-vector}
\end{align}
Similarly, summing the $\vvs{\qbar}(t)$ yields the  MGF  vector 
\begin{align}
\Ebig{e^{s\xv(t)}}=\sum_{\qbar\in\Qcal} \Ebig{e^{s\xv(t)}\delta_{\qbar,q(t)}}
=\sum_{\qbar\in\Qcal} \vvs{\qbar}(t).\eqnlabel{MGF-vector}
\end{align}
\end{subequations}

\NEW{This decomposition of the moments $\E{\xv^m(t)}$ and the MGF $\E{e^{s\xv(t)}}$ will be useful because the discrete-state probabilities $\vvm[0]{\qbar}(t)$ and the weighted conditional expectations  $\vvm{\qbar}(t)$, and $\vvs{\qbar}(t)$  are deterministic functions of time $t$ that obey a system of first-order ordinary differential equations. However, some definitions are still needed to write these equations. 
Recalling that a transition $l\in\Lcal$ is from state $q_l$ to state $q'_l$, let}
\begin{subequations}\eqnlabel{Lcalqbar}
\begin{align}
\Lcal_{\qbar}&=\set{l\in\Lcal: q_l=\qbar},\\
\Lcal'_{\qbar}&=\set{l\in\Lcal: q'_l=\qbar}
\end{align}
\end{subequations}
denote the respective sets of outgoing and incoming transitions for state $\qbar\in\Qcal$.
In addition, for each transition mapping $\Amat_l$, define a diagonal companion  matrix
$\hat{\Amat}_l$ such that
\begin{equation}\eqnlabel{hatAmat}
\cvec{\hat{\Amat}_l}_{i,j}=\begin{cases}
1 & i=j, \cvec{\Amat_l}_j=\zerov^\top,\\
0 & \ow.
\end{cases}
\end{equation} 
The nonzero entries of $\hat{\Amat}_l$ mark the zero columns of $\Amat_l$. These definitions
enable the following lemma.
\begin{lemma}\label{lem:pi-vv-derivs}  
For state $\qbar\in \Qcal$ in an  AoI SHS $(\Qcal,\onev[],\Acal)$, 
\begin{subequations}\eqnlabel{vv-derivs}
\begin{align}
\vvmdot[0]{\qbar}(t)&=\sum_{l\in\Lcal'_{\qbar}}\laml
\vvm[0]{\ql}(t)- \vvm[0]{\qbar}(t)\sum_{l\in\Lcal_{\qbar}}\laml,
\eqnlabel{pi-derivs}\\
\vvmdot{\qbar}(t)&=
m\vvm[m-1]{\qbar}(t)+\sum_{l\in\Lcal'_{\qbar}}\laml \vvm{\ql}(t)\Amat_l
\nn
&\qquad-\vvm{\qbar}(t)\sum_{l\in\Lcal_{\qbar}}\laml,
\eqnlabel{vvm-derivs}\\
\vvsdot{\qbar}(t)&=
s\vvs{\qbar}(t)+\sum_{l\in\Lcal'_{\qbar}}\laml \bracket{\vvs{\ql}(t)\Amat_l+\vvs[0]{\ql}(t)\hat{\Amat}_l}\nn
&\qquad-\vvs{\qbar}(t)\sum_{l\in\Lcal_{\qbar}}\laml.
\eqnlabel{vvs-derivs}
\end{align}
\end{subequations}
\end{lemma}
\NEW{The derivation of  Lemma~\ref{lem:pi-vv-derivs} is deferred to Section~\ref{sec:SHS}.}
From a given initial condition at time $t=0$,  \eqnref{pi-derivs}  computes the temporal evolution of the state probabilities $\vvm[0]{\qbar}(t)$.  Given $\vvm[0]{\qbar}(t)$, \eqnref{vvm-derivs} computes the conditional first moments $\vvm[1]{\qbar}(t)$. Moreover, repeating this process for $m=2,3,\ldots$ enables the successive calculation of the conditional moments $\vvm[m]{\qbar}(t)$ from $\vvm[m-1]{\qbar}(t)$  to whatever order $m$ is desired.  Each step of this process must solve a system of first order linear differential equations with $\nlen|\Qcal|$ variables. In the transform domain, the situation in \eqnref{vvs-derivs} is even simpler; given $\vvs[0]{\qbar}(t)$, a single set of $\nlen|\Qcal|$ differential equations defines $\vvs{\qbar}(t)$.

Note that Lemma~\ref{lem:pi-vv-derivs} is the foundation for the rest of this paper. In Section~\ref{sec:stationary}, the fixed points of the Lemma~\ref{lem:pi-vv-derivs} differential equations,  are used to find the stationary moments and the stationary MGF of the AoI. Section~\ref{sec:line-network} then derives the stationary moments and MGF for a line network of preemptive servers. Next, Section~\ref{sec:sampling} generalizes these results to  networks in which nodes ``sample'' the age processes of their neighbors.

\section{Stationary Moments and the Stationary MGF}\label{sec:stationary}
While using Lemma~\ref{lem:pi-vv-derivs} to calculate  age moment trajectories may be of interest, the primary value of the lemma is in deriving stationary moments of the AoI.  However,  the stationary age moments are meaningful only if  the Markov chain $q(t)$ is ergodic.  Under this ergodicity assumption, the state probabilities
$\vvm[0]{\qbar}(t)$ in \eqnref{pi-derivs}
always  converge to unique stationary probabilities $\vvmbar[0]{q}$
satisfying
\begin{align}\eqnlabel{vvmbar-stateprobs}
\vvmbar[0]{\qbar}\sum_{l\in\Lcal_{\qbar}}\laml&=\sum_{l\in\Lcal'_{\qbar}}\laml
\vvmbar[0]{\ql},
\qquad
\sum_{\qbar\in\Qcal}\vvmbar[0]{\qbar}
=\onev[].
\end{align}
Notably, \eqnref{pi-derivs} in Lemma~\ref{lem:pi-vv-derivs} shows that this convergence  of state probabilities is disconnected from the evolution of the age process. This is as expected since the age  process $\xv(t)$ is a measurements process that does not influence the evolution of the discrete state of the network.

Going forward, it is assumed that the  Markov chain $q(t)$ is ergodic and that the state probabilities $\vvm[0]{\qbar}(t)$ are given by the stationary probabilities $\vvmbar[0]{\qbar}$. Under these ergodicity and stationarity assumptions,  \eqnref{vvm-derivs}   in Lemma~\ref{lem:pi-vv-derivs}  reduces for $m=1$ to the system of first order differential equations 
\begin{align}
\vvmdot[1]{\qbar}(t)&=
\vvmbar[0]{\qbar}+\sum_{\mathclap{l\in\Lcal'_{\qbar}}}\laml \vvm[1]{\ql}(t)\Amat_l
-\vvm[1]{\qbar}(t)\sum_{l\in\Lcal_{\qbar}}\laml
\eqnlabel{vv-derivs-pibar}
\end{align}
in the vectors $\smallset{\vvm[1]{\qbar}(t):\qbar\in\Qcal}$.
While the differential equations \eqnref{vv-derivs-pibar}  hold for any set of reset maps $\smallset{\Amat_l}$, the system of equations may or may not be stable. Stability depends on the specific collection of reset maps.\footnote{For example, if 
each $\Amat_l$ maps $x_1'=x_1$, then $x_1(t)=t$ simply tracks the passage of time and $\vm{q1}(t)$ grows without bound for all states $q$.}
When \eqnref{vv-derivs-pibar} is stable, 
each $\vvmdot[1]{\qbar}(t)\to\zerov$ and  $\vvm[1]{\qbar}(t)
\to\vvmbar[1]{\qbar}$ as $t\goes\infty$, such that
\begin{align}
\vvmbar[1]{\qbar}\sum_{l\in\Lcal_{\qbar}}\laml&=
\vvmbar[0]{\qbar}+\sum_{\mathclap{l\in\Lcal'_{\qbar}}}\laml \vvmbar[1]{\ql}\Amat_l,\quad \qbar\in\Qcal.
\eqnlabel{vvmbar1-pibar}
\end{align}
In this stable case, it  follows from \eqnref{x-decomposition} that the average age is
\begin{align}\eqnlabel{EXVsum}
\E{\xv}&=\limty{t}\E{\xv(t)}\nn 
&=\limty{t}\sum_{\qbar\in\Qcal} \E{\xv(t)\delta_{\qbar,q(t)}}
=\sum_{\qbar\in\Qcal} \vvmbar[1]{\qbar}.
\end{align}

This approach can be extended to higher order age moments and the age MGF by setting the derivatives  $\vvmdot[m]{\qbar}(t)$ and $\vvsdot{\qbar}(t)$ in Lemma~\ref{lem:pi-vv-derivs}  to zero and solving for the limiting values $\vvmbar[m]{\qbar}$ and $\vvsbar{\qbar}$.  
This leads to the following result.
 \begin{theorem}\thmlabel{age-moments} If the discrete-state Markov chain $q(t)$ is ergodic with stationary distribution $\vvmbar[0]{\qbar}>\zerov$ for all $\qbar\in\Qcal$ and \eqnref{vvmbar1-pibar} has a non-negative solution $\smallset{\vvmbar[1]{\qbar}:\qbar\in\Qcal}$, then:
 \begin{itemize}
 \item[(a)] For all $\qbar\in\Qcal$, $\vvm[m]{\qbar}(t)$ converges to $\vvmbar[m]{\qbar}$ satisfying 
 \begin{subequations}
 \begin{align}
\vvmbar{\qbar}\sum_{l\in\Lcal_{\qbar}}\laml&=
m\vvmbar[m-1]{\qbar}+\sum_{l\in\Lcal'_{\qbar}}\laml \vvmbar{\ql}\Amat_l,
\eqnlabel{vvmbar-fixedpt}
\end{align}
 and $\E{\xv^m(t)}$ converges to the stationary $m$th moment  age vector
 \begin{align}
 \E{\xv^m}&=\sum_{\qbar\in\Qcal} \vvmbar{\qbar}.
 \end{align}
 \end{subequations}
  \item[(b)] There exists $s_0>0$ such that  for all $s<s_0$, $\vvs{\qbar}(t)$, $\qbar\in\Qcal$,  converges to $\vvsbar{\qbar}$ satisfying 
 \begin{subequations}
 \begin{align}\eqnlabel{vvsbar-fixedpt}
 \!\!\vvsbar{\qbar}\sum_{l\in\Lcal_{\qbar}}\laml&\!=\!
s\vvsbar{\qbar}\!+\!\sum_{l\in\Lcal'_{\qbar}}\!\laml \!\bracket{\vvsbar{\ql}\Amat_l+\vvsbar[0]{\ql}\hat{\Amat}_l}
  \end{align}
 and the MGF $\E{e^{s\xv(t)}}$ converges to the stationary vector
 \begin{equation}
 \Ebig{e^{s\xv}}=\sum_{\qbar\in\Qcal}\vvsbar{\qbar}.
  \end{equation}
  \end{subequations}
  \end{itemize}
   \end{theorem}
 \NEW{\Thmref{age-moments} is demonstrated now with an example.}
{\BLUE
\subsection{Example: The M/M/1/1 queue with abandonment}
\label{sec:MM11abandon}
This section  analyzes the AoI at the output of an M/M/1/1 queue in which updates arrive at rate $\lambda$, are served at rate $\mu$, and an update-in-service is abandoned (i.e. discarded without completing service) at rate $\alpha$.  
Furthermore, when there is an update in service, new arrivals are blocked and cleared. The discrete state $q(t)\in\Qcal=\set{0,1}$ is the queue occupancy. A Markov chain%
\footnote{In an SHS Markov chain,  there may be multiple transitions $l$ and $l'$ from state $i$ to state $j$; the transitions have different maps $\Amat_l$ and $\Amat_{l'}$. Furthermore, the SHS can include self-transitions in which the discrete state is unchanged but a reset occurs in the continuous state.\label{footnote:SHS-MC}}
for $q(t)$ is shown in Figure~\ref{fig:MM11abandon} with transitions labeled by the index $l$. Each transition $l$ is described in Table~\ref{tab:MM11abandon}.  The continuous state is $\xv(t)=\rvec{x_1(t)&x_2(t)}$, where $x_1(t)$ is the age of a monitor that sees updates that go into service and $x_2(t)$ is the age of the destination monitor that sees updates that complete service.

The goal here is to find the limiting MGF $\E{e^{sx_2}}=\limty{t}\E{e^{sx_2(t)}}$. The first step is to use 
\eqnref{vvmbar-stateprobs} to derive the  equations
$\vvmbar[0]{0}\lambda = \vvmbar[0]{1}(\mu+\alpha)$ and
$\vvmbar[0]{0}+\vvmbar[0]{1}=\onev[]$
for the Markov chain stationary probabilities. This yields the stationary probabilities
\begin{align}
\vvmbar[0]{0}=\frac{\mu+\alpha}{\lambda+\mu +\alpha}\onev[],
\qquad
\vvmbar[0]{1}=\frac{\lambda}{\lambda+\mu +\alpha}\onev[].
\end{align}

\begin{table}[t]
\caption{Table of transitions for the SHS Markov chain in Figure~\ref{fig:MM11abandon}.}
\begin{displaymath}
\setlength\arraycolsep{2pt}
\begin{array}{cccccc}
l & \laml  &q_l\to q'_l & \xv\Amat_l &\Amat_l &\hat{\Amat}_l
\\\hline
1  &\lambda &0 \to 1	& \rvec{\xz&x_2} &
\begin{bsmallmatrix} 0 & 0\\ 0 & 1\end{bsmallmatrix}&
\begin{bsmallmatrix} 1 & 0\\ 0 & 0\end{bsmallmatrix}\\
2  &\mu &1\to 0	& \rvec{x_1&x_1} &  
\begin{bsmallmatrix} 1 & 1\\ 0 & 0\end{bsmallmatrix}&
\begin{bsmallmatrix} 0 & 0\\ 0 & 0\end{bsmallmatrix}\\
3  &\alpha	&1
\to 0 & \rvec{x_1&x_2} &
\begin{bsmallmatrix} 1 & 0\\ 0 & 1\end{bsmallmatrix}&
\begin{bsmallmatrix} 0 & 0\\ 0 & 0\end{bsmallmatrix}
\end{array}
\end{displaymath}
\label{tab:MM11abandon}
\vspace{-5mm}
\end{table}

Now \eqnref{vvsbar-fixedpt} is used to write
\begin{subequations}\eqnlabel{MM11vvsbar1}
\begin{align}
\vvsbar{0}\lambda &= s\vvsbar{0}+\mu\vvsbar{1}\Amat_2 
+\alpha\vvsbar{1}\Amat_3\eqnlabel{MM11vvsbar1a},\\
\vvsbar{1}(\mu+\alpha) &= s\vvsbar{1}+\lambda\paren{\vvsbar{0}\Amat_1 
+\vvsbar[0]{0}\hat{\Amat}_1}.
\end{align}
\end{subequations}
From Table~\ref{tab:MM11abandon}, these equations simplify to 
\begin{subequations}\eqnlabel{MM1vvsbar2}
\begin{align}
\vvsbar{0}(\lambda-s)&=\vvsbar{1}
\begin{bmatrix}\mu+\alpha & \mu\\ 0 &\alpha\end{bmatrix}\!,\\
\vvsbar{1}(\mu+\alpha-s)&=\vvsbar{0}
\begin{bmatrix} 0 &0\\ 0 & \lambda\end{bmatrix}
+\vvsbar[0]{0}
\begin{bmatrix} \lambda &0\\ 0 & 0\end{bmatrix}\!.
\end{align}
\end{subequations}

For compactness of notation, let $\beta=\mu+\alpha$ and let $\gamma=\lambda+\mu+\alpha$. With some algebra, \eqnref{MM1vvsbar2} implies
\begin{subequations}\begin{align}
\vvsbar{0} &=\frac{\lambda\beta}{(\lambda-s)(\beta-s)\gamma}
\begin{bmatrix}\beta & \dfrac{\mu[\lambda\beta-\gamma s+s^2]}{\lambda\mu-\gamma s+s^2}\end{bmatrix},\\
\vvsbar{1} &=\frac{\lambda\beta}{(\beta-s)\gamma}
\begin{bmatrix} 1 & \dfrac{\lambda\mu}{\lambda\mu-\gamma s + s^2}\end{bmatrix}.
\end{align}\end{subequations}
By \Thmref{age-moments}(b), the MGF $\E{e^{s\xv(t)}}$ converges to the stationary vector
\begin{IEEEeqnarray}{rCl}
 \E{e^{s\xv}}&=&\vvsbar{0}+\vvsbar{1}\nn
 &=&\frac{\lambda\beta\begin{bmatrix}
 \gamma-s &\dfrac{\mu[\lambda\gamma-(\lambda+\gamma)s+s^2]}{\lambda\mu-\gamma s+s^2}
 \end{bmatrix}}{\gamma(\lambda-s)(\beta-s)}.
  \end{IEEEeqnarray}
The average age at the destination monitor is
\begin{align}
\E{x_2} &=\eval{\deriv{}{\E{e^{sx_2 }}}{s}}_{s=0}\nn &=\frac{1}{\lambda}+\frac{1}{\mu}+\frac{\lambda}{(\mu+\alpha)(\lambda+\mu+\alpha)}+\frac{\alpha}{\lambda\mu}.\eqnlabel{AoI-MM1abandon}
\end{align}
Note that it follows from \eqnref{AoI-MM1abandon} that a nonzero abandonment rate $\alpha$ will reduce average AoI when $\lambda/\mu$ is sufficiently large.\footnote{This observation is analogous to the conclusion in \cite{Kam-KNWE-IT2018deadline} that using deadlines to discard updates-in-service also can reduce average AoI.}
\begin{figure}
\centering
\begin{tikzpicture}[->,>=stealth', auto, semithick, node distance=2.75cm]
\tikzstyle{every state}=[fill=none,thick,scale=1]
\node[state]    (0)                     {$0$};
\node[state]    (1)[right of=0]   {$1$};
\path
(0) 	edge[bend left=20,above]     node{$1$}     	(1)
(1) 	edge[bend left=20,below]     node{$2$}     	(0)
edge[bend left=60,below] node {$3$} (0);
\end{tikzpicture}
\caption{The SHS Markov chain for the M/M/1/1 queue with abandonment. The transition rates and reset maps for links $l\in\set{1,2,3}$ are shown in Table~\ref{tab:MM11abandon}.}
\label{fig:MM11abandon}
\end{figure}
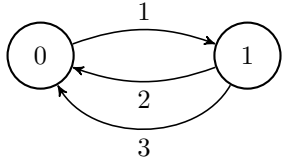

\subsection{Matrix Formulation of the Age Moments}
\label{sec:matrix-moments}
\BLACK
Lemma~\ref{lem:pi-vv-derivs} and \Thmref{age-moments} show how the ``probability balance'' in \eqnref{vvmbar-stateprobs} for the Markov chain is similar to the  ``age balance''  in \eqnref{vvmbar1-pibar}.  This makes \Thmref{age-moments} convenient for deriving the age moments of systems with a small number of states, \NEW{such as in the  M/M/1/1 example of the previous section}. 
However, the proof of \Thmref{age-moments},  \NEW{as well as computational methods for larger problems},  require a matrix representation of the systems of differential equations and their fixed points in terms of the long row vectors
\begin{subequations}\eqnlabel{long-row-vector}
\begin{align}
\vvm{}(t)&=\rvec{\vvm{0}(t) & \cdots&\vvm{\qmax}(t)},
\eqnlabel{long-row-vector-m}\\
\vvs{}(t)&=\rvec{\vvs{0}(t) & \cdots&\vvs{\qmax}(t)}.
\eqnlabel{long-row-vector-s}
\end{align}
\end{subequations}
These definitions enable the development  and proof of \Thmref{age-moments-matrix}, the matrix version of \Thmref{age-moments}(a) for the age moments,   and \Thmref{vectorMGF},  the matrix version of \Thmref{age-moments}(b) for the age MGF.

Starting with the differential equations \eqnref{vv-derivs}, define the departure rate from state $\qbar$ as 
\begin{equation}\eqnlabel{dqbar-defn}
d_{\qbar}=\sum_{l\in\Lcal_{\qbar}}\laml
\end{equation}
and the $\nlen|\Qcal|\times \nlen|\Qcal|$ diagonal matrix
\begin{align}\eqnlabel{Ddefn}
\Dmat&=\diag{d_0\Imat_{\nlen},\ldots,d_{\qmax}\Imat_{\nlen}}.
\end{align}
Defining 
\begin{align}\eqnlabel{Lcalij}
\Lcal_{i,j}&=\set{l\in\Lcal: q_l=i,q'_l=j},\quad  i,j\in\Qcal,
\end{align}
as the set of SHS transitions from state $i$ to state $j$,  let $\Rmat$ and $\hat{\Rmat}$  denote block matrices such that for  $i,j\in\Qcal$, blocks $i,j$  of $\Rmat$ and $\hat{\Rmat}$ are given by
\begin{subequations}\eqnlabel{Rmat-hat-defns}
\begin{align}\eqnlabel{Rmat-defn}
\Rmat_{i,j} &=  \sum_{l\in\Lcal_{i,j}}\laml \Amat_l,\\
\hat{\Rmat}_{i,j} &=  \sum_{l\in\Lcal_{i,j}}\laml \hat{\Amat}_l.\eqnlabel{Rmathat-defn}
\end{align}
\end{subequations}

With the observation that the set of transitions into state $\qbar$ is $\Lcal'_{\qbar}=\cup_{i}\Lcal_{i,\qbar}$,   \eqnref{vvm-derivs} becomes, for all $\qbar\in\Qcal$,
\begin{align}
\vvmdot{\qbar}(t)&=
m\vvm[m-1]{\qbar}(t)\nn
&\qquad+\sum_i\sum_{\mathclap{l\in\Lcal_{i,\qbar}}}\laml \vvm{\ql}(t)\Amat_l
-d_{\qbar}\vvm{\qbar}(t).\eqnlabel{vv-derivs-pibar2}
\end{align}
With the substitution $\qbar=j$ and the observation that $\ql=i$ for all $l\in\Lcal_{i,j}$,  it follows from \eqnref{Rmat-defn} and \eqnref{vv-derivs-pibar2} that for all $j\in\Qcal$,
\begin{align}
\!\vvmdot{j}(t)&=
m\vvm[m-1]{j}(t)\!+\!\sum_i\vvm{i}(t) \sum_{\mathclap{l\in\Lcal_{i,j}}}\laml\Amat_l
-d_{j}\vvm{j}(t)\nn
&=
m\vvm[m-1]{j}(t)\!+\!\sum_i\vvm{i}(t)\Rmat_{i,j}-d_{j}\vvm{j}(t).\eqnlabel{vv-derivs-pibar3}
\end{align}
It follows from \eqnref{long-row-vector-m} and \eqnref{Ddefn}
that \eqnref{vv-derivs-pibar3} can be written in vector form as
\begin{align}\eqnlabel{vvde-matrix}
\vvmdot{}(t) =m\vvm[m-1]{}(t)+\vvm{}(t)(\Rmat-\Dmat).
\end{align}
By stationarity of $q(t)$, $\vvm[0]{}(t)=\vvmbar[0]{}$ and thus for $m=1$, 
\begin{align}\eqnlabel{vvde-matrix1}
\vvmdot[1]{}(t) =\vvmbar[0]{}+\vvm[1]{}(t)(\Rmat-\Dmat).
\end{align}

Furthermore, setting $\vvmdot[1]{}(t)=\zerov$ and solving for $\vvm[1]{}(t)=\vvmbar[1]{}$ yields 
\begin{align}\eqnlabel{DBR}
\vvmbar[1]{}\Dmat&=\vvmbar[0]{}+\vvmbar[1]{}\Rmat.
\end{align}
Note that \eqnref{vvde-matrix1} and \eqnref{DBR} are just restatements of  \eqnref{vv-derivs-pibar} and \eqnref{vvmbar1-pibar}, but   in a matrix form that provides a straightforward characterization of the fixed points of 
\eqnref{vvde-matrix1}.
\begin{lemma}\label{stable-eigenvalues}
If  $\vvmbar[0]{}>0$ and there exists a non-negative solution $\vvmbar[1]{}$ for \eqnref{DBR}, then all eigenvalues of $\Rmat-\Dmat$ have strictly negative real parts.
\end{lemma}
Lemma~\ref{stable-eigenvalues} is a minor variation on the Perron-Frobenius theorem; the proof  in the Appendix follows from non-negativity of $\Dmat$ and $\Rmat$.  Lemma~\ref{stable-eigenvalues} says that the existence of a non-negative solution $\vvmbar[1]{}$ for \eqnref{DBR} implies that the differential equation \eqnref{vvde-matrix1} for $\vvm[1]{}(t)$ is stable and that $\limty{t}\vvm[1]{}(t)= \vvmbar[1]{}$ such that
 \begin{equation}\eqnlabel{vvmbar1}
 \vvmbar[1]{}=\vvmbar[0]{}(\Dmat-\Rmat)^{-1}.
 \end{equation}
 In the following,  $\vvmbar[1]{}$ is called the stationary first moment of the age process. Similarly, $\vvmbar[m]{}$ will denote the stationary $m$th moment. Note that Lemma~\ref{stable-eigenvalues} requires  the finite state Markov chain $q(t)$ to have no transient states; all components of the stationary probability vector $\vvmbar[0]{}$ must be strictly positive. This condition is carried forward in the next theorem.
 \begin{theorem}\thmlabel{age-moments-matrix} If the discrete-state Markov chain $q(t)$ is ergodic with stationary distribution $\vvmbar[0]{}>\zerov$ and  \eqnref{DBR} has a non-negative solution $\vvmbar[1]{}$,
 then 
 $\vvm{}(t)$
 converges to 
 \begin{subequations}\eqnlabel{mth-moment}
 \begin{align}
 \vvmbar[m]{}=\rvec{\vvmbar{0}\ \cdots\ \vvmbar{\qmax}}
 =m!\,\vvmbar[0]{}\bracket{(\Dmat-\Rmat)^{-1}}^m,
 \end{align}
 and $\E{\xv^m(t)}$ converges to the stationary $m$th moment 
 \begin{align}
 \E{\xv^m}&=\sum_{q=0}^{\qmax} \vvmbar{q}.
 \end{align}
 \end{subequations}
 \end{theorem}
 Proof by induction appears in the Appendix. The proof follows almost directly from Lemma~\ref{stable-eigenvalues}. Note that \Thmref{age-moments-matrix} is just a restatement of \Thmref{age-moments}(a) in a matrix form that enables an explicit closed form solution for $\vvmbar[m]{}$. Also note  for $m=1$ that \Thmref{age-moments-matrix} is a restatement of \cite[Theorem~4]{Yates-Kaul-IT2019}  using matrix notation. 

\subsection{Matrix Formulation of the Age MGF}\label{sec:matrix-MGF}
This section follows similar steps for the MGF to construct a matrix version of \Thmref{age-moments}(b). Recalling $\Lcal'_{\qbar}=\cup_{i}\Lcal_{i,\qbar}$ is the set of transitions into state $\qbar$,   \eqnref{vvs-derivs} becomes
\begin{align}
\vvsdot{\qbar}(t)&=
s\vvs{\qbar}(t)+\sum_i\sum_{\mathclap{l\in\Lcal_{i,\qbar}}}\laml \bracket{\vvs{\ql}(t)\Amat_l+\vvs[0]{\ql}(t)\hat{\Amat}_l}\nn
&\qquad-d_{\qbar}\vvs{\qbar}(t).\eqnlabel{vvs-derivs2}
\end{align}
With the substitution $\qbar=j$ and the observation that $\ql=i$ for all $l\in\Lcal_{i,j}$,  \eqnref{Rmat-hat-defns} and \eqnref{vvs-derivs2} imply that
\begin{align}
\vvsdot{j}(t)&=s\vvs{j}(t)+\sum_i\bracket{\vvs{i}(t)\Rmat_{i,j}
+\vvs[0]{i}(t)\hat{\Rmat}_{i,j}}\nn
&\qquad-d_{j}\vvs{j}(t).\eqnlabel{vvs-derivs3}
\end{align}
It follows from \eqnref{long-row-vector-s}, \eqnref{Ddefn} and \eqnref{Rmat-hat-defns} that \eqnref{vvs-derivs3} can be written in vector form as
\begin{align}\eqnlabel{vvsde-matrix}
\vvsdot{}(t) =\vvs{}(t)[\Rmat-\Dmat+s\Imat]+\vvs[0]{}(t)\hat{\Rmat}.
\end{align}
Under the stationarity assumption, the state probability vector is $\vvs[0]{}(t)=\vvmbar[0]{}$. Moreover, if there exists a non-negative solution $\vvmbar[1]{}$ for \eqnref{DBR}, then the eigenvalues of $\Rmat-\Dmat$ have strictly negative real parts. This implies there exists $s_0>0$ such that for all $s<s_0$ the eigenvalues of $\Rmat-\Dmat+s\Imat$ have strictly negative real parts 
and $\vvs{}(t)$ will converge to the fixed point given by $\vvsdot{}(t)=0$. This observation is formalized in the following claim, which is a restatement of \Thmref{age-moments}(b) in matrix form.
 \begin{theorem}\thmlabel{vectorMGF} If the discrete-state Markov chain $q(t)$ is ergodic with stationary distribution $\vvmbar[0]{}>\zerov$
 and there exists a stationary first moment $\vvmbar[1]{}\ge\zerov[]$ satisfying \eqnref{DBR}, then 
  there exists $s_0>0$ such that  for all $s<s_0$, $\vvs{}(t)$ converges to the stationary MGF vector
 \begin{subequations}
 \begin{equation}\eqnlabel{vvsbar-final}
 \vvsbar{}=\rvec{\vvsbar{0} &\cdots&\vvsbar{\qmax}}=\vvsbar[0]{}\hat{\Rmat}(\Dmat-\Rmat-s\Imat)^{-1}
 \end{equation}
 and the MGF $\E{e^{s\xv(t)}}$ converges to the stationary vector
 \begin{equation}
 \E{e^{s\xv}}=\sum_{q=0}^{\qmax}\vvsbar{q}.
  \end{equation}
\end{subequations}
\end{theorem}
As one would expect, the stationary MGF is sufficient to rederive the stationary moments found in \Thmref{age-moments-matrix}. For example, rewriting \eqnref{vvsbar-final} yields
  $\vvsbar{}(\Dmat-\Rmat-s\Imat)=\vvsbar[0]{}\hat{\Rmat}$.
 Taking the derivative of both sides yields
 \begin{equation}\eqnlabel{vvs-derivs1}
 \deriv{}{\vvsbar{}}{s}(\Dmat-\Rmat-s\Imat)-\vvsbar{}=\zerov[].
 \end{equation}
 Evaluating at $s=0$ yields the \Thmref{age-moments-matrix} claim
 \begin{equation}
 \eval{\deriv{}{\vvsbar{}}{s}}_{s=0}=\vvsbar[0]{}(\Dmat-\Rmat)^{-1}=\eval{\vvmbar{}}_{m=1}.
 \end{equation}
In the same way, successive derivatives of \eqnref{vvs-derivs1} will yield $\vvmbar{}$ for $m>1$ and the corresponding higher order moments $\E{\xv^m}$ in \eqnref{mth-moment}.

While the age moments of \Thmref{age-moments}(a) and \Thmref{age-moments-matrix} are indeed a simple consequence of the MGF, it is worth re-emphasizing the pivotal role of the first moment $\vvmbar[1]{}$ of the age. The existence of the first moment, i.e. the average age, guarantees the existence of, and convergence to, the stationary MGF  vector of the age process.

\section{Preemptive line networks}\label{sec:line-network}
One can now apply \Thmref{vectorMGF} to the last-come-first-served preemptive line network depicted in Figure~\ref{fig:line-network}. In this system, source node $0$ generates fresh updates as a rate $\mu_0$ Poisson process.  These updates pass through nodes $1,2,\ldots,n-1$ to sink node $n$ such that each node $i$, $i<n$,
is a rate $\mu_i$ $\cdot$/M/1/1 server supporting preemption in service. An update departing node $i-1$ immediately goes into service at node $i$ and any preempted update at node $i$ is discarded. 
\BLUE
Thus there is no queueing and each node $i$ is either idle or serving an update. Nevertheless, the process of delivering timely updates to the sink is complex for the following reasons: 
\begin{itemize}
\item The line network is lossy as updates are discarded when they are preempted;  the departure process at each node $i>0$  is not memoryless.  
\item Updates arriving at nodes $i>1$ are not fresh; instead they are aged by their passage  through prior nodes. The age of an arriving update may be correlated with its service times at preceding nodes. That is, the interarrival time of an update  may be correlated with its age.
\item The line network has $2^n$ states, since a network state must indicate whether each node is idle or busy. 
\item Updates that reach the monitor benefit from a survivor bias; each was lucky enough in its service times to avoid being preempted. 
\end{itemize}

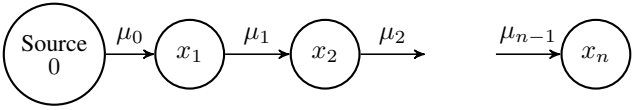
\begin{figure}[t]
\BLACK
\centering
\begin{tikzpicture}[->, >=stealth', auto, semithick, node distance=1.8cm]
\tikzstyle{every state}=[fill=none,minimum size=26pt,
draw=black,thick,text=black,scale=1]
\node[state]    (0)                     {\small \shortstack{Source\\$0$}};
\node[state]    (1)[right of=0]   {$x_1$};
\node[state]    (2)[right of=1]   {$x_2$};
\node[state,draw=none] (H1)[right of=2] {};
\node[state] (n)[right of=H1]{$x_{n}$};
\path
(0) 	edge [above]  node {$\mu_0$}    	(1)
(1)	edge[above]     node{$\mu_1$}     	(2)
(2)     edge[above] node {$\mu_2$}  (H1)  
(H1) edge[above] node {$\mu_{n-1}$} (n); 
\end{tikzpicture}
\caption{In this $n$-node  line network, $x_i$ is the age of a monitor that sees updates that go into service at node $i$. This figure admits two models: (1)  Each node $i$ is s rate $\mu_i$ $\cdot$/M/1/1 queue.  supporting preemption in service. An update departing node $i-1$ immediately goes into service at node $i$ and any preempted update at node $i$ is discarded. (2) Node $i$ forwards its current state update as a rate $\mu_i$ Poisson process to node $i+1$.}
\label{fig:line-network}
\end{figure}

Despite these issues, an initial SHS analysis \cite{Yates-aoi2018}  showed that the average  age is additive from hop to hop.
\BLACK
Specifically,  the expected age $\E{x_{n}(t)}$ at monitor $n$ was shown to converge to the stationary expected value\footnote{A discrete-time version of this result was first recognized in \cite[Theorem~1]{Talak-KM-allerton2017}.}
\begin{equation}\eqnlabel{line-network-average-age}
\E{X_{n}} = \frac{1}{\mu_0} +\frac{1}{\mu_1}+\cdots+\frac{1}{\mu_{n-1}}.
\end{equation}
\BLUE
This analysis avoided a combinatorial explosion in the SHS analysis associated with tracking the  idle/busy state of each node by employing  the method of ``fake updates'' \cite{Yates-Kaul-IT2019}. 

To describe SHS with fake updates, let $\xv(t)=\rvec{x_1(t)&\cdots&x_n(t)}$, where $x_i(t)$ is the age of a monitor that sees updates arriving at node $i$, denote the continuous state.
When an update departs node $i$ at time $t$, a ``fake update'' is created and put into service at node $i$, with the same timestamp (and age $x_i(t)$) as the update that just departed. If a new update from node $i-1$ arrives at node $i$, it preempts the fake update and the fake update causes no delay to the arriving update.  If the fake update does complete service at node $i$, it will go into service at node $i+1$, but it will have the same age as the update (whether real or fake) that it will preempt. Hence the evolution of
age vector $\xv(t)$ is the same as it would be with just real 
updates.\footnote{Note that the method of fake updates depends on preemptive service. For example, fake updates fail in a line network of the $\cdot$/M/1/1 servers supporting abandonment of service analyzed in Section~\ref{sec:MM11abandon}.  In such a network, the idle/busy state of each server must be tracked to determine whether an arriving update goes into service or is discarded.} 


Under the fake update model, each node is perpetually busy serving updates at rate $\mu_i$ for delivery to the next node. Hence, there is no need to track the idle/busy at each node. 
Moreover, each node $i$ can be viewed as offering its current update as a rate $\mu_i$ process to node $i+1$; this represents a second interpretation of the network in 
Figure~\ref{fig:line-network}. In this alternate model, a node $i$ represents a monitor that forwards its current update to node $i+1$ as a rate $\mu_i$ Poisson point process in which the transmission/delivery of an update occurs instantaneously. This idealized network will have the same  age process $\xv(t)$ as the actual M/M/1/1 line network with update packets that are preempted and servers that go idle.

\begin{table}[t]
\caption{Table of transitions for the SHS Markov chain in Figure~\ref{fig:MC-line-network}.}\label{tab:line-network}
\begin{displaymath}
\setlength\arraycolsep{2pt}
\begin{array}{ccl}
l & \laml & \xv\Amat_l 
\\\hline
0  &\mu_0 	& \rvec{\xz&x_2&x_3&\cdots& x_{n-1} & x_{n}}\\
1  &\mu_1 	& \rvec{x_1&x_1&x_3&\cdots& x_{n-1} & x_{n}} \\
2  &\mu_2	& \rvec{x_1&x_2&x_2& \cdots& x_{n-1} & x_{n}} \\
 \vdots & \vdots & \vdots\\
n-1  &\mu_{n-1} 	& \rvec{x_1&x_2&x_3&\cdots& x_{n-1}&x_{n-1}}
\end{array}
\end{displaymath}
\end{table}
\begin{figure}
\centering
\begin{tikzpicture}[>=stealth', auto, semithick, node distance=3cm]
\tikzstyle{every state}=[fill=none,draw=black,thick,text=black,scale=1]
\node[state]    (0)                     {$0$};
\path
(0) 	
 edge[loop right,right]  node {$n-1$} (0)
(0) edge[loop above, above] node {$2$} (0)
(0) edge[loop left, left] node {$0$} (0);
\draw[->] (0) to [out=150,in=120,looseness=8] node {$1$} (0);
\draw ($(0)+(0.6,0.72)$) node {$\ddots$};
\end{tikzpicture}
\caption{The SHS Markov chain for the line network with $n$ nodes.
The transition rates and transition/reset maps for links $l=0,\ldots,n$ are shown in Table~\ref{tab:line-network}.}
\label{fig:MC-line-network}
\end{figure}
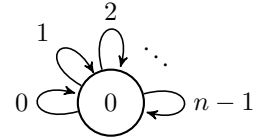
\subsection{Age Moments on the Line Network}\label{sec:line-moments}
In this section, \Thmref{age-moments-matrix} is applied to the idealized line network to enable evaluation of the age moments.
\BLACK
The SHS has the trivial discrete state space $\Qcal=\set{0}$ and stationary probabilities $\vvsbar[0]{}=\vvsbar[0]{0}=\onev[]$. From \eqnref{mth-moment-vector}, the $m$th moment of the age at time $t$  is 
\begin{align}
\vvm{}(t)&=\vvm{0}(t)=\E{\xv^m(t)}.
\end{align}
The state transitions are shown in Table~\ref{tab:line-network}. Note that:
\begin{itemize}
\item Transition $l=0$ marks the arrival of a fresh update at node $1$. In the continuous state $\xv$, this transition sets $x'_1=0$ but all other  components of $\xv$ are unchanged.
\item In a transition $l\in\set{1,\ldots,n-1}$, an update is sent from  node $l$ to node $l+1$.  At node $l+1$, the age is reset to $x'_{l+1}=x_l$.  At all other nodes, the age is unchanged. 
\end{itemize}
All transitions are trivially $0\to0$ self transitions and the total rate of transitions out of state $0$ is $d_0=\sum_{i=0}^{n-1}\mu_i$. Thus 
\begin{align}\eqnlabel{Dmat-line}
\Dmat&=d_0\Imat
\end{align} 
in \eqnref{Ddefn}. From Table~\ref{tab:line-network},  the $n\times n$ transition matrices $\Amat_l$  can be inferred and   $\Rmat=\Rmat_{00}$ given in \eqnref{Rmat-defn} can be constructed.
By following these steps and applying \Thmref{age-moments-matrix}, the following claim is verified in the Appendix.
\begin{theorem} \thmlabel{line-network-moments}
The $n$-node line network has stationary $m$th moment age vector
\begin{align}
\vvmbar{} &=\limty{t}\begin{bmatrix}
\E{x_1^m(t)}&\E{x_{2}^m(t)}&\cdots & \E{x_{n}^m(t)}
\end{bmatrix}\nn
&=m!\,\onev[] \begin{bmatrix}
\frac{1}{\mu_0} &\frac{1}{\mu_0}
& \cdots & \frac{1}{\mu_0}\\
& \frac{1}{\mu_1}
&\dots&
\frac{1}{\mu_1}\\
&         & \ddots &\vdots\\
&          &            &  \frac{1}{\mu_{n-1}}
\end{bmatrix}^m.
\end{align}
\end{theorem}

\subsection{Age MGF on the Line Network}\label{sec:line-MGF}
For the line network, using \Thmref{vectorMGF} to find the  age MGF vector requires $\Dmat$ in \eqnref{Dmat-line} and $\Rmat=\Rmat_{00}$ in \eqnref{line-network-Rmat} from the derivation of \Thmref{line-network-moments}, in order to construct  $\hat{\Rmat}$ as specified by \eqnref{Rmathat-defn}. Since $\Qcal=\set{0}$, it follows from \eqnref{MGF-vector} that the MGF vector is 
\begin{align}
\Ebig{e^{s\xv(t)}}&=\vvs{}(t)=\vvs{0}(t).
\end{align}
By applying \Thmref{vectorMGF}, the following claim is verified in the Appendix.
\begin{theorem} \thmlabel{line-network-MGF}
The $n$-node line network has stationary MGF
\begin{align}
\vvsbar{} &=\limty{t}\begin{bmatrix}
\E{e^{sx_1(t)}}&\E{e^{sx_{2}(t)}}&\cdots & \E{e^{sx_{n}(t)}}
\end{bmatrix}\nn
&=\begin{bmatrix}
\dfrac{\mu_0}{\mu_0-s} &\displaystyle\prod_{i=0}^1 \dfrac{\mu_i}{\mu_i-s} &\cdots&  \displaystyle\prod_{i=0}^{n-1}\dfrac{\mu_i}{\mu_i-s}
\end{bmatrix}.
\end{align}
\end{theorem}
The interpretation of \Thmref{line-network-MGF} is that $X_k$, the stationary age at node $k$ has distribution 
\begin{equation}\eqnlabel{indepZsum}
X_k\sim Z_0\oplus\cdots \oplus Z_{k-1},
\end{equation} 
where the $Z_i$ are independent exponential $(\mu_i)$ random variables. Note that \Thmref{line-network-MGF} generalizes the sum result \eqnref{line-network-average-age} for the average age originally derived  in \cite{Yates-aoi2018}.

\section{Sampling Stationary Age Processes}\label{sec:sampling}
\NEW{This section  examines age in networks (described by an AoI SHS) which have been running for a sufficiently long time to ensure that the  Markov chain $q(t)$ has stationary probabilities and that the MGF vector $\vvs{}(t)$ has converged to the fixed point $\vvsbar{}$. In the following definition,  a network that meets these conditions is said to have a stationary age process.}
\begin{definition} The AoI SHS $(\Qcal,\onev[],\Acal)$ is a {\em stationary age process} if the Markov chain $q(t)$ has stationary probabilities $\vvmbar[0]{}>\zerov[]$, there exists non-negative first moment $\vvmbar[1]{}$ satisfying \eqnref{DBR}, and $\vvs{}(t)=\vvsbar{}$.
\end{definition}

Convergence to a stationary MGF vector in \Thmref{vectorMGF} indicates that each component $x_j(t)$ of the age vector $\xv(t)$ has converged to a stationary distribution $\cdf{X_j}{\cdot}$. 
These conditions hold by initializing the system with $\vvs{}(0)=\vvsbar{}$. Practically, this is a weak assumption since \Thmref{vectorMGF} tells us that the existence of the first moment $\vvmbar[1]{}$ implies  exponential convergence of $\vvs{}(t)$ to the fixed point $\vvsbar{}$.

For the stationary age process $\xv(t)$ of the line network in Section~\ref{sec:line-network}, it follows from \eqnref{indepZsum} that the distribution of the age $X_{i+1}$ at node $i$ is described by\footnote{Recall that $\sim$ denotes equality in distribution and that $\oplus$ is the sum of random variables under the product distribution.}
\begin{equation}
X_{i+1}\sim X_{i}\oplus Z_{i},\eqnlabel{indep-sum}
\end{equation}
where $Z_i$ is an exponential $(\mu_i)$ random variable independent of $X_i$. 
From the original perspective of update packets and preemptive queues, this is a counterintuitive observation as the preemption and discarding \NEW{of packets} along the line network is a complex process with memory.  However, this result will prove less surprising when  $X_{i+1}$ is viewed as the age of a monitor positioned at node $i+1$.  In particular, there is a rate $\mu_i$  Poisson point process of updates being conveyed instantaneously from the monitor at node $i$ to the monitor at node $i+1$. An arrival of this process at time $t$ resets $x_{i+1}(t)$ to $x_i(t)$. Between arrivals of these updates, the age process at node $i+1$ grows at unit rate.  In fact, this is an example of the more general situation in this definition: 
\begin{definition}
\label{def:sampling} Node $j$ is {\em sampling the update process} at node $i$ if 
\begin{enumerate}
\item[(a)] $x_i(t)$ is an age process for a monitor at node $i$ that maintains a {\em current update}, i.e. a copy of its freshest received update. 
\item[(b)]  There is a point process that marks the time instances that the current update at node $i$ is delivered to  node $j$. 
\item[(c)] When node $i$ delivers an update to $j$ at time $t'$, the age $x_j(t')$  is reset to   $x'_j(t')=x_{i}(t')$, the age of the just-received update. 
\item[(d)] In the absence of an update from node $i$, the age at node $j$ grows at unit rate. 
\end{enumerate} 
\end{definition}
Node $j$ is said to  be sampling the status update process  at node $i$ because node $j$ receives a subsequence of those state samples  delivered to node $i$. While this sampling is defined in terms of the state updates sent from $i$ to $j$, it is also convenient  from the perspective of age analysis to say node $j$ is {\em sampling the age process} at node $i$.

In the case of the Figure~\ref{fig:line-network} line network, each node $j=i+1$ is sampling the node $i$ status update process as a rate $\mu_i$ Poisson process. The inter-update times of this sampling process are iid exponential $(\mu_i)$ random variables.  The Poisson sampling on the line network can be generalized to sampling through a renewal process in which the renewal points mark the time instances that updates from $i$ are delivered to $j$. 

\subsection{Status Sampling Renewal Processes}
To analyze the 
status-sampling renewal process, suppose  at time $t$ that the most recent update delivery from node $i$ occurred at time $t-Z(t)$. \NEW{Because the updates are forwarded as a renewal process, $Z(t)$ is the time since the most recent renewal. That is, $Z(t)$ is the age (aka the backwards excess) of the renewal process.} In particular, $Z(t)$ will have a sample path like that \NEW{of $x_1(t)$} shown in Figure~\ref{fig:cloudnet}(\NEW{b}).  When this renewal process is in equilibrium and has inter-update times that are iid continuous random variables identical  to $Y$, $Z(t)$  is stationary and has PDF \cite[Theorem~5.7.4]{Gallager2013stochastic}
\begin{equation}\eqnlabel{ZPDF}
\pdf{Z}{z}=\frac{\prob{Y>z}}{\E{Y}},\qquad z\ge 0.
\end{equation}
When the status sampling process is a rate $\mu$ Poisson process,  $Y$ is an exponential $(\mu)$ random variable and it follows from \eqnref{ZPDF} that $Z$ is also exponential $(\mu)$.  Moreover, conditions (c) and (d) in Definition~\ref{def:sampling} imply that the age at node $j$ is
\begin{equation}
X_{j}(t)=X_i(t-Z(t))+Z(t).\eqnlabel{age-sampling}
\end{equation}
The next claim follows from \eqnref{age-sampling}.
\begin{theorem}\thmlabel{sampling} If node $j$ is sampling a stationary age process $X_i(t)$ at node $i$ according to an equilibrium renewal process with inter-update times identical to $Y$ and stationary renewal process age $Z$ distributed according to \eqnref{ZPDF}, then the age $X_j(t)$ at node $j$ is a stationary process identical in distribution to 
\begin{equation}
X_j\sim X_i\oplus Z.
\end{equation}
\end{theorem}
\begin{IEEEproof}(\Thmref{sampling}) It follows from \eqnref{age-sampling} and the definition of the MGF that
\begin{subequations}\eqnlabel{cond-MGFX}
\begin{align}
\Ebig{e^{sX_{j}(t)}|Z(t)=z}
&=\Ebig{e^{s(X_i(t-z)+z)}|Z(t)=z}\\
&=e^{sz}\Ebig{e^{sX_i(t-z)}|Z(t)=z}\\
&=e^{sz}\Ebig{e^{sX_i(t-z)}}\eqnlabel{Xizi-indep}\\
&=e^{sz}\Ebig{e^{sX_i(t)}}.\eqnlabel{Xi-stationary}
\end{align}
\end{subequations}
Note that  \eqnref{Xizi-indep} follows from independence of the $X_i(t)$ and $Z(t)$ processes,   and \eqnref{Xi-stationary} holds because $X_i(t)$ is stationary. Thus $\E{e^{sX_{j}(t)}|Z(t)}=e^{sZ(t)}\E{e^{sX_i(t)}}$ and averaging over $Z(t)$ yields
\begin{IEEEeqnarray}{rCl}
\Ebig{e^{sX_{j}(t)}}&=&\Ewrtbig[Z(t)]{e^{sX_{j}(t)}|Z(t)}\nn
&=&\Ewrtbig[Z(t)]{e^{sZ(t)}e^{sX_i(t)}}\nn
&=&\Ewrtbig{e^{sZ(t)}}\Ebig{e^{sX_i(t)}}.\eqnlabel{MGFX}
\end{IEEEeqnarray}
Since $X_i(t)$ and $Z(t)$ are stationary processes, they are 
identical to $X_i$ and $Z$ in distribution and thus \eqnref{indep-sum} holds.
\end{IEEEproof}

For the line network in Figure~\ref{fig:line-network}, the derivation of  the stationary age distribution in \eqnref{indep-sum} relied on the age process $\xv(t)$ being generated by an AoI SHS. However, this assumption is not required in the steps of \eqnref{cond-MGFX} and \eqnref{MGFX}. All that is needed is that $X_i(t)$ is a stationary age process that is independent of the equilibrium renewal process that controls the sampling.

This notion of sampling an update process offers a generalization of the Figure~\ref{fig:line-network} line network. Instead of memoryless updating,  node $i+1$ now samples the update process at node $i$  according to a renewal process with inter-renewal times identical to random variable $Y_i$. This network is depicted in Figure~\ref{fig:line-sampling}. When each $Y_i$ has an exponential $(\mu_i)$ distribution, the age at each network node is identical to the age in Theorems~\thmref{line-network-moments} and~\thmref{line-network-MGF} for the  line network of preemptive servers in Figure~\ref{fig:line-network}.  In general however, the network in Figure~\ref{fig:line-sampling} should not be mistaken for a queueing network as there are no easily defined customer service times.  Instead, it is an example of a {\em status sampling network} in which each node samples the updating process of each preceding node in the line.

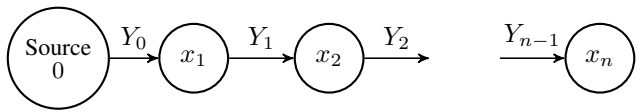
\begin{figure}
\centering
\begin{tikzpicture}[->, >=stealth', auto, semithick, node distance=1.8cm]
\tikzstyle{every state}=[fill=none,minimum size=26pt,
draw=black,thick,text=black,scale=1]
\node[state]    (0)                     {\small \shortstack{Source\\$0$}};
\node[state]    (1)[right of=0]   {$x_1$};
\node[state]    (2)[right of=1]   {$x_2$};
\node[state,draw=none] (H1)[right of=2] {};
\node[state] (n)[right of=H1]{$x_{n}$};
\path
(0) 	edge [above]  node {$Y_0$}    	(1)
(1)	edge[above]     node{$Y_1$}     	(2)
(2)     edge[above] node {$Y_2$}  (H1)  
(H1) edge[above] node {$Y_{n-1}$} (n); 
\end{tikzpicture}
\caption{The linear status sampling network with $n$ nodes. Node $i+1$ samples the update process at node $i$ at time instances that form a renewal process with renewal times identical to $Y_i$.}
\label{fig:line-sampling}
\end{figure}

In this network, status sampling  at node $i$ is an equilibrium renewal process with stationary age $Z_i$ with PDF in the form of \eqnref{ZPDF} 
and moments
\begin{align}\eqnlabel{Z-moments}
\E{Z_i}&=\frac{\E{Y_i^2}}{2\E{Y_i}},\qquad
\E{Z_i^2} =\frac{\E{Y_i^3}}{3\E{Y_i}}.
\end{align}
The age at source node $0$ is always zero and is trivially stationary. 
By \Thmref{sampling}, $X_{i+1}\sim X_i\oplus Z_i$ and stationarity of $X_i(t)$ implies stationarity  of $X_{i+1}(t)$. It follows that the stationary age at node $k$ has distribution
\begin{align}\eqnlabel{age-sum}
X_k&\sim Z_0\oplus\cdots\oplus Z_{k-1}.
\end{align}

This behavior can be demonstrated with a simulation of the status sampling network shown in Figure~\ref{fig:line-sampling} in which each sampling renewal time $Y_i$ is a continuous uniform $(0,b)$ random variable $Y$ and each $Z_i$ has PDF given by 
\eqnref{ZPDF}.  Starting at time $0$ with the age vector $\xv(0)=\zerov$, status samples forwarded by node $i$ are used to generate the age process $X_{i+1}(t)$ at node $i+1$. Specifically, at node $i$, the iid sequence $Y_{i,1},Y_{i,2},\ldots, Y_{i,m}$ of uniform $(0,b)$ update interarrival times is generated. At each time $T_{i,j}=\sum_{k=1}^j Y_{i,k}$, the age sample $X_i(T_{i,j})$ is forwarded to node $i+1$. This sequence of samples yields the $Z(t)$ function in \eqnref{age-sampling} with $j=i+1$, and this enables construction of the age process $X_{i+1}(t)$. This age sample path construction process is successively repeated at each node and is depicted graphically in Figure~\ref{fig:sampled-age-paths}.\footnote{In discrete time, this same sample path construction has previously appeared in \cite{Talak-KM-allerton2017}.}

At each node $i$, samples of the age process $X_i(t)$ are used to form the normalized histograms shown in Figure~\ref{fig:sample-histograms}.  
From \eqnref{age-sum}, the PDF of each $X_k$  is the $k$-fold convolution of the PDF of $Z$. In particular, when the inter-update time $Y$ has a uniform $(0,b)$ PDF, \eqnref{ZPDF} implies
\begin{align}
\pdf{X_1}{x} &= \begin{cases}
\frac{2}{b}\paren{1-\frac{x}{b}} & 0\le x\le b,\\
0 & \ow,
\end{cases}\\
\shortintertext{and}
\pdf{X_2}{x} &=\begin{cases}
\frac{4x}{b^2}\paren{1-\frac{x}{b}+\frac{x^2}{6b^2}} & 0\le x\le b,\\
\frac{2}{3b}\paren{2-\frac{x}{b}}^3 & b\le x\le 2b,\\
0 & \ow.
\end{cases}
\end{align}
In Figure~\ref{fig:sample-histograms}(a), histograms from sample paths  of $X_1(t)$ and $X_2(t)$ are shown to be in correspondence with the  PDFs $\pdf{X_1}{x}$ and $\pdf{X_2}{x}$.

\begin{figure}[t]
\centering
\includegraphics[scale=1.1]{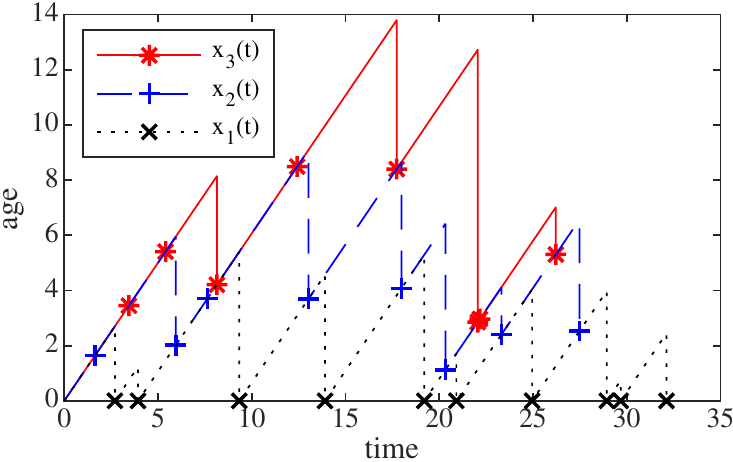}
\caption{Sample paths of age processes $x_1(t)$, $x_2(t)$ and $x_3(t)$; the markers $\color{red}*$, $\BLUE+$ and $\times$ mark when node $i$  samples the update process at the preceding node. For each node, status sampling times are chosen according to an independent renewal process. Node $1$ receives fresh updates, the age process $x_1(t)$ is reset to zero when an update is received. 
}
\label{fig:sampled-age-paths}
\end{figure}

In addition, it follows from \eqnref{Z-moments} that each $Z_i$ has moments $\E{Z}=b/3$ and $\E{Z^2}=b^2/6$. Under stationarity, each $X_k(t)$ has expected value $\E{X_k}=kb/3$ and variance $\Var{X_k}=k\Var{Z}=kb^2/18$.
As $n$ becomes large, the central limit theorem will take effect. To examine this convergence, Figure~\ref{fig:sample-histograms}(b) compares the sample histograms against the Gaussian PDFs of the same mean and variance as $X_n$ for $n=3,4,5$. Even for small $n$, it can be seen  that the Gaussian approximation is a reasonable fit. The distribution of $X_n$ is skewed leftward relative to the corresponding Gaussian; but this is not surprising inasmuch as the uniform PDF of $Y$ implies the PDF of each $Z$ is skewed to the left.

\section{Stochastic Hybrid Systems for AoI Analysis}\label{sec:SHS}
\NEW{This section  casts the AoI-tracking SHS introduced in Section~\ref{sec:newSHSintro} in the general formalism of SHS.} 
While there are many SHS variations \cite{Teel-SS-2014stability}, this work follows the model and notation in \cite{Hespanha-2006modelling}. Here it is shown that the age-tracking SHS of Definition~\ref{def:SHS-AoI}  is a special special case of the general SHS model in \cite{Hespanha-2006modelling}. \NEW{Formulating age-tracking in this general model does  pay a price in terms of  cumbersome notation, but it avoids reinventing the wheel. The direct application of Dynkin's formula to the age tracking SHS will yield Lemma~\ref{lem:pi-vv-derivs}, which is the basis for Theorems~\thmref{age-moments}, \thmref{age-moments-matrix}, and 
\thmref{vectorMGF}.
 Furthermore, this development may open a door to extensions of the age-tracking SHS model.}
 
 \begin{figure}[t]
\begin{tabular}{c}
\includegraphics[scale=1]{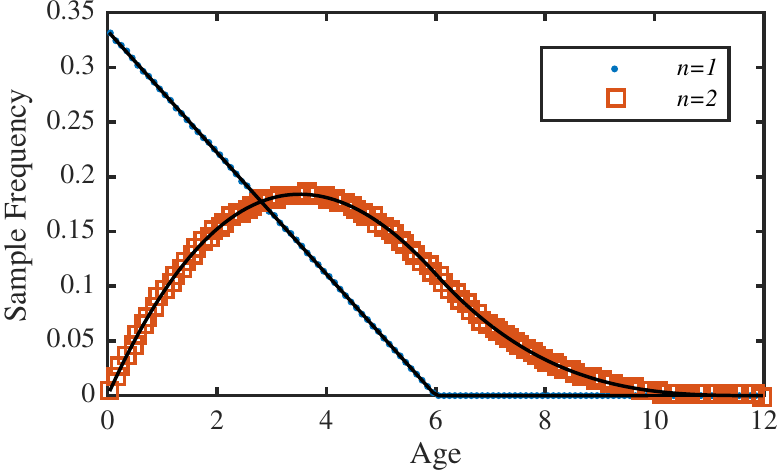}\\
{\bf (a)}  $n=1,2$\\[5mm]
\includegraphics[scale=1]{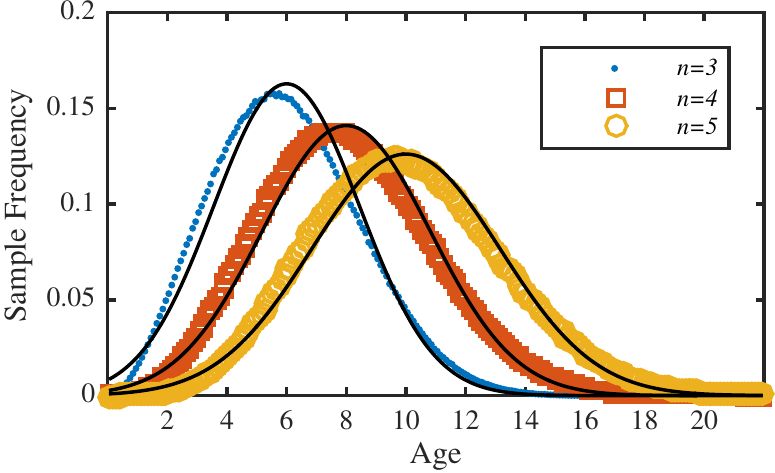}\\
{\bf (b)} $n=3,4,5$
\end{tabular}
\caption{Sample frequency histograms from sample paths of $X_n(t)$ for $n\in\set{1,\ldots,5}$. Each $X_n(t)$ is a stationary process with $\E{X_n}=2n$ and $\var{X_n}=2n$. (a) For $n=1,2$, normalized histgrams are shown to match the  corresponding PDFs $\pdf{X_1}{x}$ and $\pdf{X_2}{x}$. (b) For $n=3,4,5$, each histogram is compared with the approximating  Gaussian PDF of the same mean and variance.}
\label{fig:sample-histograms}
\end{figure}
 
 As noted in Section~\ref{sec:newSHSintro}, the SHS state is partitioned into a discrete component  $q(t)\in\Qcal=\set{0,1,\ldots,\qmax}$ that evolves as a point process and a continuous component  $\xv(t)=\rowvec{x_{\first}(t) & \cdots &x_n(t)}\in\R^{\nlen}$.  Given the discrete set  $\Qcal$ and the $k$-vector $\zv(t)$ of independent Brownian motion processes,  an SHS is defined by a stochastic differential equation  
 \begin{equation}\eqnlabel{SHSde-general}
 \dot{\xv}=f(q,\xv,t) + g(q,\xv,t)\dot{\zv}
 \end{equation}
 for mappings $f:\Qcal\times \R^{\nlen}\times[0,\infty)\to\R^{\nlen}$ and $g: \Qcal\times \R^{\nlen}\times [0,\infty)\to\R^{\nlen\times k}$, and a set of transitions $\Lcal$ such that each $l\in \Lcal$ defines  a discrete transition/reset map 
 \begin{subequations}
 \eqnlabel{SHStrans}
 \begin{IEEEeqnarray}{c}
 (q',\xv') =\phi_l(q,\xv,t),\quad \phi_l\!:\!\Qcal\!\times\! \R^{\nlen}\!\times\! [0,\infty)\to\Qcal\!\times\!\R^{\nlen},\eqnlabel{SHSjump}\IEEEeqnarraynumspace
 \end{IEEEeqnarray}
 with transition intensity
\begin{IEEEeqnarray}{c}
 \laml(q,\xv,t),\quad  \laml\!:\!\Qcal\!\times\!\R^{\nlen}\!\times\! [0,\infty)\to [0,\infty).
 \eqnlabel{SHSlaml}\IEEEeqnarraynumspace
 \end{IEEEeqnarray}
 \end{subequations}
When the system is in discrete state $q$, $\xv(t)$ evolves according to \eqnref{SHSde-general}; but  in a discrete transition from $q$ to $q'$, the continuous state can make a discontinuous jump from $\xv$ to $\xv'$, as specified by \eqnref{SHSjump}. 
Associated with each transition $l$ is a counting process $N_l(t)$ that counts the number of occurrences of transition $l$ in the interval $[0,t]$. The probability that $N_l$ jumps in the interval $(t,t+dt]$ is $\laml(q(t),\xv(t),t)\,dt$.     

\subsection{AoI tracking as an SHS}\label{sec:SHS-AoI} 
 In terms of the general SHS model given by \eqnref{SHSde-general} and
\eqnref{SHStrans}, the AoI tracking system introduced in Section~\ref{sec:newSHSintro} 
is a simple \NEW{time invariant} SHS in which
\begin{subequations}\eqnlabel{simplified}
\begin{align}
f(q,\xv,t)&=\onev[],\eqnlabel{simplified-f}\\
g(q,\xv,t)&=0,\eqnlabel{simplified-g}\\
\laml(q,\xv,t)&=\laml\dq{\ql},\eqnlabel{simplified-lam}\\
\phi_l(q,\xv,t) &=(q'_l,\xv\Amat_l).\eqnlabel{simplified-map}
\end{align}
\end{subequations}
Applying the conditions of \eqnref{simplified} to the general SHS framework in \eqnref{SHStrans} yields the age-tracking SHS of Definition~\ref{def:SHS-AoI}. In particular, \NEW{note that $q$ in \eqnref{simplified} represents a sample value of the discrete state $q(t)$. Thus,} \eqnref{simplified-f} and \eqnref{simplified-g} imply that the continuous state evolution \eqnref{SHSde-general} in each discrete state $q(t) = q$ is  simply $\dot{\xv}(t)=\onev[]$, as specified in \eqnref{SHSde}.
In \eqnref{simplified-lam}, the Kronecker delta function $\dq{q_l}$ ensures that transition $l$ occurs only \NEW{if $q(t)=q=q_l$.} Thus, $q(t)$ is a continuous-time finite-state Markov chain
%
such that each transition $l$ is a directed edge $(q_l,q'_l)$ with fixed transition rate $\laml$ while $q(t) = q_l$.   Furthermore, for each transition $l$, \eqnref{simplified-map} ensures that the transition reset map is the  linear mapping 
$\xv'=\xv\Amat_l$. 

\subsection{SHS test functions}\label{sec:testfunctions}
Because of the generality and power of the SHS model, characterization of the $q(t)$ and $\xv(t)$ processes can be complicated and often intractable. The approach in \cite{Hespanha-2006modelling} is to define test functions $\psi(q,\xv,t)$ whose expected values $\E{\psi(q(t),\xv(t),t)}$ are performance measures of interest that can be evaluated as functions of time; see \cite{Hespanha-2006modelling}, \cite{Hespanha-course}, and the survey \cite{Teel-SS-2014stability} for additional background.

\BLUE
Since the simplified SHS defined in \eqnref{simplified} is time invariant, this work will employ time invariant test functions $\psi(q,\xv)$. Specifically, for each age process $x_j(t)$, the $m$th moment $\E{x_j^m(t)}$ and  the MGF $\E{e^{sx_j(t)}}$ are  tracked in each state $\qbar\in\Qcal$ using the families of test functions  $\smallset{\psim{\qbar, j}(q,\xv):\qbar\in\Qcal, j\in\range{n}}$ and $\smallset{\psis{\qbar, j}(q,\xv):\qbar\in\Qcal,j\in\range{n}}$ such that  
\begin{subequations}\eqnlabel{testfcn-S:defn}
\begin{align}
\psim{\qbar, j}(q,\xv) &= x_j^m\dq{\qbar},\quad m=0,1,2,\ldots,\\
\shortintertext{and}
\psis{\qbar, j}(q,\xv) &= e^{s x_j}\dq{\qbar}.
\end{align}
\end{subequations}
These test functions have expected values  
\begin{subequations}\eqnlabel{orig-vqi-defn}
\begin{align}
\Ebig{\psim{\qbar, j}(q(t),\xv(t))}&=\Ebig{x_j^m(t)\dqt{\qbar}}=\vm{\qbar, j}(t),\eqnlabel{orig-vqi0}\\
\Ebig{\psis{\qbar, j}(q(t),\xv(t))}&=\Ebig{e^{sx_j(t)}\dqt{\qbar}}= \vs{\qbar, j}(t)
\end{align}
\end{subequations}
that are the weighted conditional expectations defined in \Eqnref{vqi-defn}.

The objective here is to use the SHS framework to derive the system of differential equations  in Lemma~\ref{lem:pi-vv-derivs} for $\vvm{\qbar}(t)$ and $\vvs{\qbar}(t)$ defined in \eqnref{vv-defn}. To do so, the SHS mapping $\psi\to L\psi$ known as the extended generator is applied to every test function $\psi(q,\xv)$, from either family $\smallset{\psim{\qbar, j}(q,\xv)}$ or  $\smallset{\psis{\qbar, j}(q,\xv)}$, The extended generator $L\psi$ is simply a function whose expected value is the expected rate of change of the test function $\psi$.
\BLACK
Specifically, a test function $\psi(q(t),\xv(t))$  has an extended generator $(L\psi)(q(t),\xv(t))$ that  satisfies Dynkin's formula
\begin{align}
\eqnlabel{dynkins}
\deriv{}{\E{\psi(q(t),\xv(t))}}{t}&=\E{(L\psi)(q(t),\xv(t))}.
\end{align}
\NEW{For each test function $\psi(q,\xv)$, \eqnref{dynkins} yields a differential equation for $\E{\psi(q(t),\xv(t))}$.}

From \cite[Theorem~1]{Hespanha-2006modelling}, it follows from the conditions \eqnref{simplified} and the time invariance of either version of $\psi(q,\xv)$ in \eqnref{testfcn-S:defn} that 
the extended generator of a piecewise linear SHS is given by
\begin{align}\eqnlabel{Lpsi-defn}
(L\psi)(q,\xv)
&= \frac{\partial\psi(q,\xv)}{\partial \xv}\onev[]^\top\nn &\quad+\sum_{l\in\Lcal} \bracket{\psi(\phi_l(q,x))-\psi(q,x)}\laml(q),
\end{align}
where the row vector $\partial\psi(q,\xv)/\partial\xv$ denotes the gradient. When $\psi(q,\xv)=\psim{\qbar, j}(q,\xv)$,
\begin{subequations}\eqnlabel{partial-derivs}
\begin{align}
\frac{\partial\psim{\qbar, j}(q,\xv)}{\partial\xv}
&=mx_j^{m-1}\dq{\qbar}\ev_{j}=m\psim[m-1]{\qbar, j}(q,\xv)\ev_j\\
\shortintertext{and when $\psi(q,\xv)=\psis{\qbar, j}(q,\xv)$,}
\frac{\partial\psis{\qbar, j}(q,\xv)}{\partial\xv}
&=se^{sx_j}\dq{\qbar}\ev_{j}=s\psis{\qbar, j}(q,\xv)\ev_j.
\end{align}
\end{subequations}

With the definitions of $\Lcal'_{\qbar}$, $\Lcal_{\qbar}$ and $\hat{\Amat}_l$ in \eqnref{Lcalqbar} and \eqnref{hatAmat},  applying \eqnref{Lpsi-defn} and \eqnref{partial-derivs} to \eqnref{dynkins}, yields Lemma~\ref{lem:pi-vv-derivs}, the system of first order ordinary differential equations for $\vvm[0]{\qbar}(t)$, $\vvm{\qbar}(t)$, and $\vvs{\qbar}(t)$. 
The algebraic steps, i.e. the proof of Lemma~\ref{lem:pi-vv-derivs},  appear in the Appendix.

\BLACK
\section{Conclusion}
\label{sec:conclusions}
This work has employed an age of information metric to analyze the timeliness of a source delivering status updates through a network to a monitor.  \NEW{For any network described by a finite-state continuous-time Markov chain, a stochastic hybrid system was employed  to identify a system of ordinary linear differential equations that describe the temporal evolution of the moments  and MGF of an age process vector.  Solving for the fixed point of these differential equations enables} evaluation of all stationary moments of the age as well as the MGF of the age.  To the best of my knowledge, the MGF vector results are the first explicit age distribution results for updates traversing multihop routes in networks. 

In the example of the line network with preemptive memoryless servers, a generalization of the prior SHS analysis \cite{Yates-aoi2018} showed that the age at a node is distributed according to a sum of independent renewal process age random variables. In the case of preemptive memoryless servers, these renewal process age random variables were also exponentially distributed. However, 
this observation was generalized in Section~\ref{sec:sampling}  by showing it holds for all  equilibrium renewal processes with continuous inter-update distributions and stationary age processes. The key to this generalization is the concept of sampling an update process.

The insights from these simple observations suggest there is still considerable progress to be made in characterizing age of information in systems and networks.  
\appendix
\begin{IEEEproof} (Lemma~\ref{lem:pi-vv-derivs})
\label{sec:lem:pi-vv-derivs}
In the following, $m$ is a strictly positive integer. Equations \eqnref{Lpsi-defn} and \eqnref{partial-derivs} enable calculation of  $\Lpsim[0]{\qbar, j}$, $ \Lpsim{\qbar, j}$ and $\Lpsis{\qbar, j}$ for each $j\in\range{n}$ and $\qbar\in\Qcal$:
\begin{subequations}\eqnlabel{Lpsim-S-defn}
\begin{align}
\Lpsim[0]{\qbar, j}(q,\xv)&=\Lamm[0]{\qbar, j}(q,\xv),\eqnlabel{Lpsim-S-defn0}\\
\Lpsim{\qbar, j}(q,\xv)
&=m\psim[m-1]{\qbar, j}(q,\xv) + \Lamm{\qbar, j}(q,\xv), 
\eqnlabel{Lpsim-S-defn-m}\\
\Lpsis{\qbar, j}(q,\xv) &= s\psis{\qbar, j}(q,\xv) + \Lams{\qbar, j}(q,\xv),
\eqnlabel{Lpsim-S-defn-s}
\end{align}
\end{subequations}
where 
\begin{subequations}\eqnlabel{Lammiqbar}
\begin{IEEEeqnarray}{rCl}
\Lamm[0]{\qbar, j}(q,\xv) &=& \sum_{l\in\Lcal}
\bracket{\psim[0]{\qbar, j}(\phi_l(q,\xv))
-\psim[0]{\qbar, j}(q,\xv)}\laml(q),\IEEEeqnarraynumspace\\
\Lamm{\qbar, j}(q,\xv)&=&\sum_{l\in\Lcal}
\bracket{\psim{\qbar, j}(\phi_l(q,\xv))
-\psim{\qbar, j}(q,\xv)}\laml(q),\IEEEeqnarraynumspace\\
\Lams{\qbar, j}(q,\xv)&=&\sum_{l\in\Lcal}
\bracket{\psis{\qbar, j}(\phi_l(q,\xv))
-\psis{\qbar, j}(q,\xv)}\laml(q).\IEEEeqnarraynumspace
\end{IEEEeqnarray}
\end{subequations}

When $\phi_l(q,\xv)=(\ql',\xv\Amat_l)$,
\begin{subequations}\eqnlabel{psim1}
\begin{align}
\psim[0]{\qbar, j}(\phi_l(q,\xv))&=\psim[0]{\qbar, j}(\ql',\xv\Amat_l) = \dq[\ql']{\qbar},\\
\psim{\qbar, j}(\phi_l(q,\xv))&=\psim{\qbar, j}(\ql',\xv\Amat_l)\nn
&=\cvec{\xv\Amat_l}_j^m\dq[\ql']{\qbar}
=\cvec{\xv^m\Amat_l}_j\dq[\ql']{\qbar},
\eqnlabel{xm-swap}\\
\psis{\qbar, j}(\phi_l(q,\xv))&=\psis{\qbar, j}(\ql',\xv\Amat_l)
=e^{s\cvec{\xv\Amat_l}_j}\dq[\ql']{\qbar}.\eqnlabel{esx-swap}
\end{align}
\end{subequations}
In \eqnref{xm-swap},  $\cvec{\xv\Amat_l}_j^m=\cvec{\xv^m\Amat_l}_j$ because  $\Amat_l$ is a binary matrix such that each column has no more than one nonzero entry. 
Since $\laml(q)=\laml\dq{\ql}$, it follows from \eqnref{Lammiqbar} and \eqnref{psim1} that
\begin{subequations}\eqnlabel{Lamm-S-defn}
\begin{align}
\Lamm[0]{\qbar, j}(q,\xv)&=\sum_{l\in\Lcal}\laml\bracket{\dq[\ql']{\qbar}-\dq{\qbar}}\dq{\ql},\\
\Lamm{\qbar, j}(q,\xv)&=\sum_{l\in\Lcal}\laml\bracket{[\xv^m\Amat_l]_j\dq[\ql']{\qbar}-x_j^m\dq{\qbar}}\dq{\ql},\\
\Lams{\qbar, j}(q,\xv)&=\sum_{l\in\Lcal}\laml\bracket{e^{s[\xv\Amat_l]_j}\dq[\ql']{\qbar}-e^{s x_j}\dq{\qbar}}\dq{\ql}.
\end{align}
\end{subequations}

Note that
 \begin{subequations}\eqnlabel{kronecker-sift}
\begin{align}
\dq[\ql']{\qbar}\dq{\ql}&=\begin{cases}
\dq{\ql} & l \in \Lcal'_{\qbar},\\
0 & \ow,
\end{cases}\\
\dq{\qbar}\dq{\ql} &=\begin{cases}
\dq{\qbar} & l\in\Lcal_{\qbar},\\
0 & \ow.
\end{cases}
\end{align}
\end{subequations}
It follows from \eqnref{Lcalqbar}, \eqnref{Lamm-S-defn} and \eqnref{kronecker-sift} that
\begin{subequations}\eqnlabel{Lammlast}
\begin{IEEEeqnarray}{rCl}
\Lamm[0]{\qbar, j}(q,\xv)&=&\sum_{l\in\Lcal'_{\qbar}}\laml\dq{\ql}-\dq{\qbar}\sum_{l\in\Lcal_{\qbar}}\laml,\eqnlabel{Lammlast-pi}\IEEEeqnarraynumspace\\
\Lamm{\qbar, j}(q,\xv)&=&\sum_{l\in\Lcal'_{\qbar}}\laml {\cvec{\xv^m\Amat_l}_j}
\dq{\ql}-x_j^m\dq{\qbar}\sum_{l\in\Lcal_{\qbar}}\laml,
\eqnlabel{Lammlast-j}\IEEEeqnarraynumspace\\
\Lams{\qbar, j}(q,\xv)&=&\sum_{l\in\Lcal'_{\qbar}}\laml e^{s\cvec{\xv\Amat_l}_j}
\dq{\ql}-e^{sx_j}\dq{\qbar}\sum_{l\in\Lcal_{\qbar}}\laml.\IEEEeqnarraynumspace
\eqnlabel{Lamslast-j}
\end{IEEEeqnarray}
\end{subequations}

With $\psi(q,\xv)=\psim[0]{\qbar, j}(q,\xv)$,  \eqnref{vqi0} at $m=0$, 
 \eqnref{dynkins}, \eqnref{Lpsim-S-defn0} and \eqnref{Lammlast-pi} imply for all $\qbar\in\Qcal$ that
\begin{align}
\vmdot[0]{\qbar, j}(t)&=\Ebig{\Lpsim[0]{\qbar, j}(q(t),\xv(t))}\nn
&=\Ebig{\Lamm[0]{\qbar, j}(q(t),\xv(t))}\nn
&=\EBig{\sum_{l\in\Lcal'_{\qbar}}\laml\dqt{\ql}-\dqt{\qbar}\sum_{l\in\Lcal_{\qbar}}\laml}\nn
&=
\sum_{l\in\Lcal'_{\qbar}}\laml
\vm[0]{\ql}(t)- \vm[0]{\qbar}(t)\sum_{l\in\Lcal_{\qbar}}\laml.\eqnlabel{statprobs}
\end{align}
Gathering the equations in \eqnref{statprobs}, and rewriting in terms of the vector of duplicated state probabilities $\vvm[0]{\qbar}(t)$, yields \eqnref{pi-derivs} in Lemma~\ref{lem:pi-vv-derivs}.

For $j\in \range{n}$ with $\psi(q,\xv)=\psim{\qbar, j}(q,\xv)$,  \eqnref{vqi-defn}, \eqnref{dynkins} and \eqnref{Lpsim-S-defn-m}  imply
\begin{align}
\vmdot{\qbar, j}(t)&=\Ebig{\Lpsim{\qbar, j}(q(t),\xv(t))}\nn
&=\Ebig{m\psim[m-1]{\qbar, j}(q(t),\xv(t))}
+\Ebig{\Lamm{\qbar, j}(q(t),\xv(t))}\nn
&= m\vm[m-1]{\qbar, j}(t)+\Ebig{\Lamm{\qbar, j}(q(t),\xv(t))}.
\eqnlabel{viqde1}
\end{align}
From \eqnref{Lammlast-j}, it follows that
\begin{align}
&\Ebig{\Lamm{\qbar, j}(q(t),\xv(t))}\nn
&=\sum_{l\in\Lcal'_{\qbar}}\laml 
\bracket{\E{\xv^m(t)\dqt{\ql}}\Amat_l}_j-\E{x^m_j(t)\dqt{\qbar}}\sum_{l\in\Lcal_{\qbar}}\laml\nn
&=\sum_{l\in\Lcal'_{\qbar}}\laml\bracket{\vvm{\ql}(t)\Amat_l}_j
-\vm{\qbar, j}(t)\sum_{l\in\Lcal_{\qbar}}\laml.\eqnlabel{Lammlast2}
\end{align}
It then follows from \eqnref{viqde1} and \eqnref{Lammlast2} that 
\begin{align}
\vmdot{\qbar, j}(t)
&=
m \vm[m-1]{\qbar, j}(t)
+\sum_{l\in\Lcal'_{\qbar}}\laml \bracket{\vvm{\ql}(t)\Amat_l}_j\nn
&\qquad-\vm{\qbar, j}(t)\sum_{l\in\Lcal_{\qbar}}\laml.\eqnlabel{vmqde2}
\end{align}
 Gathering the equations in \eqnref{vmqde2} for $j\in\range{n}$, and rewriting as row vectors, yields  \eqnref{vvm-derivs} in Lemma~\ref{lem:pi-vv-derivs}.
   
For $j\in \range{n}$ with $\psi(q,\xv)=\psis{\qbar, j}(q,\xv)$,  \eqnref{vqi-defn}, \eqnref{dynkins} and \eqnref{Lpsim-S-defn-s}  imply
\begin{align}
\vsdot{\qbar, j}(t)&=\Ebig{\Lpsis{\qbar, j}(q(t),\xv(t))}\nn
&=\Ebig{s \psis{\qbar, j}(q(t),\xv(t))}
+\Ebig{\Lams{\qbar, j}(q(t),\xv(t))}\nn
&=s\vs{\qbar, j}(t)+\Ebig{\Lams{\qbar, j}(q(t),\xv(t))}.
\eqnlabel{vsqde1}
\end{align}
From \eqnref{Lamslast-j}, 
\begin{align}
&\Ebig{\Lams{\qbar, j}(q(t),\xv(t))}\nn
&=\sum_{l\in\Lcal'_{\qbar}}\laml \Ebig{e^{s\cvec{\xv(t)\Amat_l}_j}\dqt{\ql}}-\Ebig{e^{sx_j(t)}\dqt{\qbar}}\sum_{l\in\Lcal_{\qbar}}\laml\nn 
&=\sum_{l\in\Lcal'_{\qbar}}\laml\Ebig{e^{s\cvec{\xv(t)\Amat_l}_j}\dqt{\ql}}
-\vs{\qbar, j}(t)\sum_{l\in\Lcal_{\qbar}}\laml.\eqnlabel{Lamslast2}
\end{align}
Similarly, if $[\Amat_l]_j=\ev^\top_k$, then $\cvec{\xv\Amat_l}_j=x_k$ and
\begin{subequations}\eqnlabel{esxA-cases}
\begin{align}
\Ebig{e^{s[\xv(t)\Amat_l]_j}\dqt{\ql}}&=\Ebig{e^{s x_k(t)}\dqt{\ql}}\nn
&=\vs{\ql k}(t)=\cvec{\vvs{\ql}(t)\Amat_l}_j.
\end{align}
 However, if $[\Amat_l]_j=\zerov^\top$, then 
 \begin{align}
 \Ebig{e^{s\cvec{\xv(t)\Amat_l}_j}\dqt{\ql}}&=\E{\dqt{\ql}}=\vs[0]{\ql}(t).
 \end{align}
  \end{subequations}
 With the definition of $\hat{\Amat}_l$ in \eqnref{hatAmat}, 
 the two cases in \eqnref{esxA-cases} can be written in the combination form  
 \begin{align}
 \Ebig{e^{s\cvec{\xv(t)\Amat_l}_j}\dqt{\ql}}
 &=\cvec{\vvs{\ql}(t)\Amat_l+\vvs[0]{\ql}(t)\hat{\Amat}_l}_j
 \eqnlabel{EesxA}
 \end{align}
because either $[\Amat_l]_j=\zerov^\top$ or $[\hat{\Amat}_l]_j=\zerov^\top$. It then follows from \eqnref{vsqde1}, \eqnref{Lamslast2} and \eqnref{EesxA} that 
\begin{align}
\vsdot{\qbar, j}(t)
&=
s \vs{\qbar, j}(t)
+\sum_{l\in\Lcal'_{\qbar}}\laml\cvec{\vvs{\ql}(t)\Amat_l+\vvs[0]{\ql}(t)\hat{\Amat}_l}_j\nn 
&\qquad-\vs{\qbar, j}(t)\sum_{l\in\Lcal_{\qbar}}\laml.\eqnlabel{vsqde2}
\end{align}
 To complete the proof, gathering the equations \eqnref{vsqde2} for all $j$ and rewriting as row vectors yields \eqnref{vvs-derivs} in Lemma~\ref{lem:pi-vv-derivs}.
 \end{IEEEproof}
 
 \medskip
 
 \begin{IEEEproof} (Lemma~\ref{stable-eigenvalues})
Let $\sigma=1+\max_i d_i$, then $\sigma\Imat -\Dmat$ is a strictly positive diagonal matrix.  Adding $\vvmbar[1]{}(\sigma\Imat-\Dmat)$ to both sides of \eqnref{DBR} yields
 \begin{align}\eqnlabel{DBR-uniformized}
\sigma\vvmbar[1]{}
&=\vvmbar[0]{}+\vvmbar[1]{}(\sigma\Imat+\Rmat-\Dmat).
\end{align}
Because the reset maps $\Amat_l$ are binary,  $\Rmat$ is  non-negative and  thus $\sigma\Imat+\Rmat-\Dmat$ is also non-negative. It follows that  $\sigma\Imat+\Rmat-\Dmat$ has a dominant real eigenvalue $r(\sigma)\ge0$ with an associated non-negative non-zero right eigenvector $\uv^\top$ such that $\abs{\epsilon}\le r(\sigma)$ for any other eigenvalue $\epsilon$  \cite[Exercise 1.12]{Seneta-1981non-negative}\footnote{This is a weak form of the Perron-Frobenius theorem that does not require irreducibility of the non-negative matrix.}. Right multiplying \eqnref{DBR-uniformized} by $\uv^\top$ produces
\begin{align}
\sigma\vvmbar[1]{}\uv^\top=\vvmbar[0]{}\uv^\top 
+r(\sigma)\vvmbar[1]{}\uv^\top,
\end{align}
which simplifies to
\begin{align}
[\sigma-r(\sigma)]\vvmbar[1]{}\uv^\top=\vvmbar[0]{}
\uv^\top. \eqnlabel{rs-simplified}
\end{align}
Since $\vvmbar[0]{}$ is strictly positive and $\uv^\top$ is non-negative and non-zero, $\vvmbar[0]{}\uv^\top> 0$. Non-negativity of $\vvmbar[1]{}$ implies $\vvmbar[1]{}\uv^\top\ge 0$ and it follows that $\sigma-r(\sigma)>0$.
If $\epsilon$ is an eigenvalue of $\sigma\Imat+\Rmat-\Dmat$ then $\epsilon-\sigma$ is an eigenvalue of $\Rmat-\Dmat$ with real part
$\real{\epsilon-\sigma} =\real{\epsilon}-\sigma
\le \abs{\epsilon}-\sigma
\le r(\sigma)-\sigma<0$.
\end{IEEEproof}

\medskip

\begin{IEEEproof} (\Thmref{age-moments-matrix})
Proof  is by induction. For $m=1$, it follows from Lemma~\ref{stable-eigenvalues} that the existence of the fixed point $\vvm[1]{}$ implies that $\Rmat-\Dmat$ has stable eigenvalues and that the differential equation \eqnref{vvde-matrix1} converges to $\limty{t}\vvm[1]{}(t)=\vvmbar[1]{}$.  By the induction hypothesis,
 \begin{align}
 \limty{t}\vvm[m-1]{}(t)&=\vvmbar[m-1]{}\nn
 &=(m-1)!\vvmbar[0]{}[(\Dmat-\Rmat)^{-1}]^{m-1}.
 \end{align}  
 Since $\Rmat-\Dmat$ has stable eigenvalues, it follows from  \eqnref{vvde-matrix} that $\vvmdot{}(t)\to\zerov$  and $\vvm{}(t)\to\vvmbar{}$ satisfying
 \begin{equation}
 \zerov=m\vvmbar[m-1]{}+ \vvmbar{}(\Rmat-\Dmat).
 \end{equation}
 Thus,
 \begin{equation}
 \vvmbar{}= m\vvmbar[m-1]{}(\Dmat-\Rmat)^{-1}
 =m!\vvmbar[0]{}[(\Dmat-\Rmat)^{-1}]^m.
 \end{equation}
 To complete the proof,  note that
 \begin{align}\eqnlabel{EXVm-sum}
\E{\xv^m}
&=\limty{t}\sum_{\qbar\in\Qcal} \E{\xv^m(t)\delta_{\qbar,q(t)}}
=\sum_{\qbar\in\Qcal} \vvmbar{\qbar}.
\end{align}
 \end{IEEEproof}
 
 \medskip
 
 \begin{IEEEproof}(\Thmref{line-network-moments})
Table~\ref{tab:line-network} yields the $n$ dimensional transition matrices $\Amat_l$; for example,
\begin{subequations}\eqnlabel{line-network-Amats}
\begin{align}
\Amat_0&=\diag{0,1,\ldots,1},\eqnlabel{line-network-Amat0}\\
\Amat_1&=
\begin{bmatrix}
1 & 1  \\
0 & 0      \\
   &   & 1  \\
   &   & & \ddots   \\
   &   &         &    &          1
\end{bmatrix},\\
\shortintertext{and}
\Amat_{n-1}&=
\begin{bmatrix}
1   \\
   &  1   \\
   &   & \ddots   \\
   &   &        &  1& 1\\
   &   &        &  0& 0
\end{bmatrix}.
\end{align}
\end{subequations}
With the shorthand notation $\beta_j=d_0 -\mu_j$,  \eqnref{Rmat-defn} can be used to construct
\begin{align}\eqnlabel{line-network-Rmat}
\Rmat=\Rmat_{00} &= \sum_{l=0}^{n-1}\mu_l \Amat_l
=\begin{bmatrix}
\beta_0 & \mu_1 \\
           & \beta_1&\mu_2\\
           &             & \beta_2&\ddots\\
           &             &            & \ddots    &\mu_{n-1}\\
           &             &            &        &\beta_{n-1}
           \end{bmatrix}.
\end{align}
It then follows from \eqnref{Dmat-line} that
\begin{align}\eqnlabel{line-network-D-R}
\Dmat-\Rmat&=
\begin{bmatrix}
\mu_0 & -\mu_1 \\
           & \mu_1&\ddots\\
           &             &   \ddots          &-\mu_{n-1}\\
           &             &            &\mu_{n-1}
           \end{bmatrix}
\shortintertext{and}
(\Dmat-\Rmat)^{-1}&=
\begin{bmatrix}
\frac{1}{\mu_0} &\frac{1}{\mu_0}& \cdots & \frac{1}{\mu_0}\\
& \frac{1}{\mu_1} &\dots& \frac{1}{\mu_1}\\
&         & \ddots &\vdots\\
&          &            &  \frac{1}{\mu_{n-1}}
\end{bmatrix}.
\end{align}
The claim follows from \Thmref{age-moments-matrix}.
\end{IEEEproof}
 
 \medskip
 
\begin{IEEEproof} (\Thmref{line-network-MGF})
From Table~\ref{tab:line-network}, the transition matrices $\Amat_l$ are given in \eqnref{line-network-Amats}. Similarly, $\Rmat$ and $\Dmat-\Rmat$  are given in \eqnref{line-network-Rmat} and \eqnref{line-network-D-R}.  Since  $\Amat_i$ has no all-zero columns for $i\ge1$, it follows from \eqnref{Rmathat-defn} and \eqnref{line-network-Amat0} that 
\begin{align}\eqnlabel{line-network-Rhat}
\hat{\Rmat}=\hat{\Rmat}_{00}&=\mu_0\hat{\Amat}_0=
\diag{\mu_0, 0, \cdots, 0}.
\end{align}
Defining 
$\mubar_i=\mu_i-s$,
$i\in0:n-1$,
it follows from \eqnref{line-network-D-R} that
\begin{align}
\Dmat-\Rmat-s\Imat=
\begin{bmatrix}
\mubar_0 & -\mu_1 \\
           & \mubar_1&-\mu_2\\
           &             & \mubar_2&\ddots\\
           &             &            &  \ddots   &-\mu_{n-1}\\
           &             &            &         &\mubar_{n-1}
           \end{bmatrix}.
           \end{align}
This implies
\begin{align}
\!\!\!\!(\Dmat-\Rmat-s\Imat)^{-1}\!=\!\setlength\arraycolsep{2pt}\begin{bmatrix}
\frac{1}{\mubar_0} &\frac{\mu_1}{\mubar_0\mubar_1}& \frac{\mu_1\mu_2}{\mubar_0\mubar_1\mubar_2}& \cdots & \frac{1}{\mubar_0}\prod_{i=1}^n\frac{\mu_i}{\mubar_i}\\
& \frac{1}{\mubar_1} & \frac{\mu_2}{\mubar_1\mubar_2} &\dots&
\frac{1}{\mubar_1}\prod_{i=2}^n\frac{\mu_i}{\mubar_i}\\
&         & \frac{1}{\mubar_2} &\cdots& \frac{1}{\mubar_2}\prod_{i=3}^n \frac{\mu_i}{\mubar_i}\\
&         && \ddots &\vdots\\
&          &            &  &\frac{1}{\mubar_{n-1}}
\end{bmatrix}\!\!.
\end{align}
Recalling that $\vvsbar[0]{}=\onev[]$,  \eqnref{line-network-Rhat} and \Thmref{vectorMGF} imply 
\begin{align}
\vvsbar{}&=\vvsbar[0]{}\hat{\Rmat}(\Dmat-\Rmat-s\Imat)^{-1}\nn
&=\begin{bmatrix}
\frac{\mu_0}{\mubar_0}&\frac{\mu_0\mu_1}{\mubar_0\mubar_1} &\frac{\mu_0\mu_1\mu_2}{\mubar_0\mubar_1\mubar_2}& \cdots & & \prod_{i=0}^{n-1}\frac{\mu_i}{\mubar_i}
\end{bmatrix}.
\end{align}
The claim follows from the definition of $\mubar_i$. 
\end{IEEEproof}
\begin{spacing}{1.0}
\bibliographystyle{IEEEtran}
\bibliography{AOI-2020-03}

\begin{thebibliography}{100}
\providecommand{\url}[1]{#1}
\csname url@samestyle\endcsname
\providecommand{\newblock}{\relax}
\providecommand{\bibinfo}[2]{#2}
\providecommand{\BIBentrySTDinterwordspacing}{\spaceskip=0pt\relax}
\providecommand{\BIBentryALTinterwordstretchfactor}{4}
\providecommand{\BIBentryALTinterwordspacing}{\spaceskip=\fontdimen2\font plus
\BIBentryALTinterwordstretchfactor\fontdimen3\font minus
  \fontdimen4\font\relax}
\providecommand{\BIBforeignlanguage}[2]{{%
\expandafter\ifx\csname l@#1\endcsname\relax
\typeout{** WARNING: IEEEtran.bst: No hyphenation pattern has been}%
\typeout{** loaded for the language `#1'. Using the pattern for}%
\typeout{** the default language instead.}%
\else
\language=\csname l@#1\endcsname
\fi
#2}}
\providecommand{\BIBdecl}{\relax}
\BIBdecl

\bibitem{Kaul-YG-infocom2012}
S.~Kaul, R.~Yates, and M.~Gruteser, ``Real-time status: How often should one
  update?'' in \emph{Proc. IEEE INFOCOM}, March 2012, pp. 2731--2735.

\bibitem{Sun-UBYKS-IT2017UpdateorWait}
Y.~Sun, E.~Uysal-Biyikoglu, R.~D. Yates, C.~E. Koksal, and N.~B. Shroff,
  ``Update or wait: How to keep your data fresh,'' \emph{IEEE Trans. Info.
  Theory}, vol.~63, no.~11, pp. 7492--7508, Nov. 2017.

\bibitem{Yates-isit2015}
R.~Yates, ``Lazy is timely: Status updates by an energy harvesting source,'' in
  \emph{Proc. IEEE Int'l. Symp. Info. Theory (ISIT)}, June 2015, pp.
  3008--3012.

\bibitem{Yates-Kaul-IT2019}
R.~D. Yates and S.~K. Kaul, ``The age of information: Real-time status updating
  by multiple sources,'' \emph{IEEE Trans. Info. Theory}, vol.~65, no.~3, pp.
  1807--1827, mar 2019.

\bibitem{Ross1996stochastic}
S.~M. Ross, \emph{Stochastic Processes}, 2nd~ed.\hskip 1em plus 0.5em minus
  0.4em\relax John Wiley \& Sons, 1996.

\bibitem{Gallager2013stochastic}
R.~G. Gallager, \emph{Stochastic processes: theory for applications}.\hskip 1em
  plus 0.5em minus 0.4em\relax Cambridge University Press, 2013.

\bibitem{Kaul-YG-ciss2012}
S.~Kaul, R.~Yates, and M.~Gruteser, ``Status updates through queues,'' in
  \emph{Conf. on Information Sciences and Systems (CISS)}, Mar. 2012.

\bibitem{Yates-Kaul-isit2012}
R.~Yates and S.~Kaul, ``Real-time status updating: Multiple sources,'' in
  \emph{Proc. IEEE Int'l. Symp. Info. Theory (ISIT)}, Jul. 2012.

\bibitem{Kam-KE-isit2013random}
C.~Kam, S.~Kompella, and A.~Ephremides, ``Age of information under random
  updates,'' in \emph{Proc. IEEE Int'l. Symp. Info. Theory (ISIT)}, 2013, pp.
  66--70.

\bibitem{Kam-KE-isit2014diversity}
------, ``Effect of message transmission diversity on status age,'' in
  \emph{Proc. IEEE Int'l. Symp. Info. Theory (ISIT)}, June 2014, pp.
  2411--2415.

\bibitem{Kam-KNE-IT2016diversity}
C.~Kam, S.~Kompella, G.~D. Nguyen, and A.~Ephremides, ``Effect of message
  transmission path diversity on status age,'' \emph{IEEE Trans. Info. Theory},
  vol.~62, no.~3, pp. 1360--1374, Mar. 2016.

\bibitem{Yates-isit2018}
R.~D. Yates, ``Status updates through networks of parallel servers,'' in
  \emph{Proc. IEEE Int'l. Symp. Info. Theory (ISIT)}, Jun. 2018, pp.
  2281--2285.

\bibitem{Costa-CE-isit2014}
M.~Costa, M.~Codreanu, and A.~Ephremides, ``Age of information with packet
  management,'' in \emph{Proc. IEEE Int'l. Symp. Info. Theory (ISIT)}, June
  2014, pp. 1583--1587.

\bibitem{Costa-CE-IT2016management}
------, ``On the age of information in status update systems with packet
  management,'' \emph{IEEE Trans. Info. Theory}, vol.~62, no.~4, pp.
  1897--1910, April 2016.

\bibitem{Kavitha-AS-arxiv2018}
\BIBentryALTinterwordspacing
V.~Kavitha, E.~Altman, and I.~Saha, ``Controlling packet drops to improve
  freshness of information,'' \emph{CoRR}, vol. abs/1807.09325, 2018. [Online].
  Available: \url{http://arxiv.org/abs/1807.09325}
\BIBentrySTDinterwordspacing

\bibitem{Chen-Huang-isit2016}
K.~Chen and L.~Huang, ``Age-of-information in the presence of error,'' in
  \emph{Proc. IEEE Int'l. Symp. Info. Theory (ISIT)}, 2016, pp. 2579--2584.

\bibitem{Champati-AG-aoi2018}
J.~P. Champati, H.~Al-Zubaidy, and J.~Gross, ``Statistical guarantee
  optimization for age of information for the {D/G/1} queue,'' in \emph{IEEE
  Conference on Computer Communications (INFOCOM) Workshops}, April 2018, pp.
  130--135.

\bibitem{Inoue-MTT-IT2019}
Y.~{Inoue}, H.~{Masuyama}, T.~{Takine}, and T.~{Tanaka}, ``A general formula
  for the stationary distribution of the age of information and its application
  to single-server queues,'' \emph{IEEE Transactions on Information Theory},
  vol.~65, no.~12, pp. 8305--8324, 2019.

\bibitem{Kam-KNWE-isit2016deadline}
C.~Kam, S.~Kompella, G.~D. Nguyen, J.~Wieselthier, and A.~Ephremides, ``Age of
  information with a packet deadline,'' in \emph{Proc. IEEE Int'l. Symp. Info.
  Theory (ISIT)}, 2016, pp. 2564--2568.

\bibitem{Kam-KNWE-IT2018deadline}
C.~Kam, S.~Kompella, G.~D. Nguyen, J.~E. Wieselthier, and A.~Ephremides, ``On
  the age of information with packet deadlines,'' \emph{IEEE Transactions on
  Information Theory}, vol.~64, no.~9, pp. 6419--6428, 2018.

\bibitem{Wang-FY-spawc2018}
B.~{Wang}, S.~{Feng}, and J.~{Yang}, ``To skip or to switch? {Minimizing} age
  of information under link capacity constraint,'' in \emph{2018 IEEE 19th
  International Workshop on Signal Processing Advances in Wireless
  Communications (SPAWC)}, June 2018, pp. 1--5.

\bibitem{Huang-Modiano-isit2015}
L.~Huang and E.~Modiano, ``Optimizing age-of-information in a multi-class
  queueing system,'' in \emph{Proc. IEEE Int'l. Symp. Info. Theory (ISIT)},
  Jun. 2015.

\bibitem{Pappas-GKK-icc2015}
N.~{Pappas}, J.~{Gunnarsson}, L.~{Kratz}, M.~{Kountouris}, and V.~{Angelakis},
  ``Age of information of multiple sources with queue management,'' in
  \emph{2015 IEEE International Conference on Communications (ICC)}, 2015, pp.
  5935--5940.

\bibitem{Kadota-UBSM-Allerton2016}
I.~Kadota, E.~Uysal-Biyikoglu, R.~Singh, and E.~Modiano, ``Minimizing the age
  of information in broadcast wireless networks,'' in \emph{54th Annual
  Allerton Conference on Communication, Control, and Computing (Allerton)},
  Sept 2016, pp. 844--851.

\bibitem{Kaul-Yates-isit2017}
S.~K. Kaul and R.~Yates, ``Status updates over unreliable multiaccess
  channels,'' in \emph{Proc. IEEE Int'l. Symp. Info. Theory (ISIT)}, Jun. 2017,
  pp. 331--335.

\bibitem{Najm-Telatar-aoi2018}
E.~Najm and E.~Telatar, ``Status updates in a multi-stream {M/G/1/1} preemptive
  queue,'' in \emph{IEEE Conference on Computer Communications (INFOCOM)
  Workshops}, April 2018, pp. 124--129.

\bibitem{Beytur-UB-SIU2018}
H.~B. Beytur and E.~Uysal-Bıyıkoglu, ``Minimizing age of information on
  multi-flow networks,'' in \emph{2018 26th Signal Processing and
  Communications Applications Conference (SIU)}, May 2018, pp. 1--4.

\bibitem{Jiang-KZZN-isit2018}
Z.~Jiang, B.~Krishnamachari, X.~Zheng, S.~Zhou, and Z.~Niu, ``Decentralized
  status update for age-of-information optimization in wireless multiaccess
  channels,'' in \emph{Proc. IEEE Int'l. Symp. Info. Theory (ISIT)}, June 2018,
  pp. 2276--2280.

\bibitem{Jiang-KZZN-iot2019}
Z.~{Jiang}, B.~{Krishnamachari}, X.~{Zheng}, S.~{Zhou}, and Z.~{Niu}, ``Timely
  status update in wireless uplinks: Analytical solutions with asymptotic
  optimality,'' \emph{IEEE Internet of Things Journal}, vol.~6, no.~2, pp.
  3885--3898, 2019.

\bibitem{Kostas-PEA-jcn2019}
A.~{Kosta}, N.~{Pappas}, A.~{Ephremides}, and V.~{Angelakis}, ``Age of
  information performance of multiaccess strategies with packet management,''
  \emph{Journal of Communications and Networks}, vol.~21, no.~3, pp. 244--255,
  2019.

\bibitem{Maatouk-AE-ToN2020}
A.~{Maatouk}, M.~{Assaad}, and A.~{Ephremides}, ``On the age of information in
  a {CSMA} environment,'' \emph{IEEE/ACM Transactions on Networking}, pp.
  1--14, 2020.

\bibitem{Maatouk-AE-aoi2019}
------, ``Minimizing the age of information: {NOMA} or {OMA}?'' in \emph{IEEE
  INFOCOM 2019 - IEEE Conference on Computer Communications Workshops (INFOCOM
  WKSHPS)}, 2019, pp. 102--108.

\bibitem{Sun-UBK-aoi2018}
Y.~Sun, E.~Uysal-Biyikoglu, and S.~Kompella, ``Age-optimal updates of multiple
  information flows,'' in \emph{IEEE Conference on Computer Communications
  (INFOCOM) Workshops}, April 2018, pp. 136--141.

\bibitem{Tripathi-Moharir-globecom2017}
V.~Tripathi and S.~Moharir, ``Age of information in multi-source systems,'' in
  \emph{IEEE Global Communications Conference (GLOBECOM)}, Dec 2017.

\bibitem{Jiang-KZN-itc2018}
Z.~{Jiang}, B.~{Krishnamachari}, S.~{Zhou}, and Z.~{Niu}, ``Can decentralized
  status update achieve universally near-optimal age-of-information in wireless
  multiaccess channels?'' in \emph{2018 30th International Teletraffic Congress
  (ITC 30)}, vol.~01, 2018, pp. 144--152.

\bibitem{Hsu-isit2018}
Y.-P. Hsu, ``Age of information: Whittle index for scheduling stochastic
  arrivals,'' in \emph{Proc. IEEE Int'l. Symp. Info. Theory (ISIT)}, June 2018,
  pp. 2634--2638.

\bibitem{Najm-NT-IT2020}
E.~Najm, R.~Nasser, and E.~Telatar, ``Content based status updates,''
  \emph{IEEE Transactions on Information Theory}, 2020.

\bibitem{Kaul-Yates-isit2018priority}
S.~Kaul and R.~Yates, ``Age of information: Updates with priority,'' in
  \emph{Proc. IEEE Int'l. Symp. Info. Theory (ISIT)}, Jun. 2018, pp.
  2644--2648.

\bibitem{Maatouk-AE-isit2019}
A.~{Maatouk}, M.~{Assaad}, and A.~{Ephremides}, ``Age of information with
  prioritized streams: When to buffer preempted packets?'' in \emph{2019 IEEE
  International Symposium on Information Theory (ISIT)}, 2019, pp. 325--329.

\bibitem{Huang-Qian-globecom2017}
L.~Huang and L.~P. Qian, ``Age of information for transmissions over {Markov}
  channels,'' in \emph{IEEE Global Communications Conference (GLOBECOM)}, Dec
  2017.

\bibitem{Tang-WSS-allerton2019}
H.~{Tang}, J.~{Wang}, L.~{Song}, and J.~{Song}, ``Scheduling to minimize age of
  information in multi-state time-varying networks with power constraints,'' in
  \emph{2019 57th Annual Allerton Conference on Communication, Control, and
  Computing (Allerton)}, 2019, pp. 1198--1205.

\bibitem{Parag-TC-wcnc2017realtime}
P.~Parag, A.~Taghavi, and J.~Chamberland, ``On real-time status updates over
  symbol erasure channels,'' in \emph{2017 IEEE Wireless Communications and
  Networking Conference (WCNC)}, March 2017.

\bibitem{Najm-YS-isit2017}
E.~Najm, R.~Yates, and E.~Soljanin, ``Status updates through {M/G/1/1} queues
  with {HARQ},'' in \emph{Proc. IEEE Int'l. Symp. Info. Theory (ISIT)}, Jun.
  2017, pp. 131--135.

\bibitem{Yates-NSZ-isit2017}
R.~Yates, E.~Najm, E.~Soljanin, and J.~Zhong, ``Timely updates over an erasure
  channel,'' in \emph{Proc. IEEE Int'l. Symp. Info. Theory (ISIT)}, Jun. 2017,
  pp. 316--320.

\bibitem{Ceran-GG-wcnc2018}
E.~T. Ceran, D.~G{\"u}nd{\"u}z, and A.~Gy{\"o}rgy, ``Average age of information
  with hybrid {ARQ} under a resource constraint,'' in \emph{2018 IEEE Wireless
  Communications and Networking Conference (WCNC)}, April 2018.

\bibitem{Sac-BUBD-spawc2018}
H.~Sac, T.~Bacinoglu, E.~Uysal-Biyikoglu, and G.~Durisi, ``Age-optimal channel
  coding blocklength for an {M/G/1} queue with {HARQ},'' in \emph{19th
  International Workshop on Signal Processing Advances in Wireless
  Communications (SPAWC)}, June 2018, pp. 486--490.

\bibitem{Bacinoglu-CUB-ita2015}
B.~T. Bacinoglu, E.~T. Ceran, and E.~Uysal-Biyikoglu, ``Age of information
  under energy replenishment constraints,'' in \emph{Proc. Info. Theory and
  Appl. (ITA) Workshop}, Feb. 2015, la Jolla, CA.

\bibitem{Bacinoglu-SUBM-isit2018}
B.~T. Bacinoglu, Y.~Sun, E.~Uysal-Biyikoglu, and V.~Mutlu, ``Achieving the
  age-energy tradeoff with a finite-battery energy harvesting source,'' in
  \emph{Proc. IEEE Int'l. Symp. Info. Theory (ISIT)}, Jun. 2018, pp. 876--880.

\bibitem{Wu-YW-green2018}
X.~{Wu}, J.~{Yang}, and J.~{Wu}, ``Optimal status update for age of information
  minimization with an energy harvesting source,'' \emph{IEEE Transactions on
  Green Communications and Networking}, vol.~2, no.~1, pp. 193--204, 2018.

\bibitem{Farazi-KB-aoi2018}
S.~Farazi, A.~G. Klein, and D.~R. Brown, ``Average age of information for
  status update systems with an energy harvesting server,'' in \emph{IEEE
  Conference on Computer Communications (INFOCOM) Workshops}, April 2018, pp.
  112--117.

\bibitem{Feng-Yang-isit2018}
S.~{Feng} and J.~{Yang}, ``Minimizing age of information for an energy
  harvesting source with updating failures,'' in \emph{2018 IEEE International
  Symposium on Information Theory (ISIT)}, 2018, pp. 2431--2435.

\bibitem{Baknina-Ulukus-arxiv2018coded}
A.~{Baknina} and S.~{Ulukus}, ``Coded status updates in an energy harvesting
  erasure channel,'' in \emph{52nd Annual Conference on Information Sciences
  and Systems (CISS)}, 2018, pp. 1--6.

\bibitem{Arafa-Ulukus-TW2019}
A.~{Arafa} and S.~{Ulukus}, ``Timely updates in energy harvesting two-hop
  networks: Offline and online policies,'' \emph{IEEE Transactions on Wireless
  Communications}, vol.~18, no.~8, pp. 4017--4030, 2019.

\bibitem{Arafa-YUP-IT2020}
A.~{Arafa}, J.~{Yang}, S.~{Ulukus}, and H.~V. {Poor}, ``Age-minimal
  transmission for energy harvesting sensors with finite batteries: Online
  policies,'' \emph{IEEE Transactions on Information Theory}, vol.~66, no.~1,
  pp. 534--556, 2020.

\bibitem{Talak-KM-allerton2017}
R.~Talak, S.~Karaman, and E.~Modiano, ``Minimizing age-of-information in
  multi-hop wireless networks,'' in \emph{55th Annual Allerton Conference on
  Communication, Control, and Computing}, Oct 2017, pp. 486--493.

\bibitem{He-YE-IT2018}
Q.~He, D.~Yuan, and A.~Ephremides, ``Optimal link scheduling for age
  minimization in wireless systems,'' \emph{IEEE Transactions on Information
  Theory}, vol.~64, no.~7, pp. 5381--5394, July 2018.

\bibitem{Lu-JL-mobicom2018}
\BIBentryALTinterwordspacing
N.~Lu, B.~Ji, and B.~Li, ``Age-based scheduling: Improving data freshness for
  wireless real-time traffic,'' in \emph{Proceedings of the Eighteenth ACM
  International Symposium on Mobile Ad Hoc Networking and Computing}, ser.
  Mobihoc '18.\hskip 1em plus 0.5em minus 0.4em\relax New York, NY, USA: ACM,
  2018, pp. 191--200. [Online]. Available:
  \url{http://doi.acm.org/10.1145/3209582.3209602}
\BIBentrySTDinterwordspacing

\bibitem{Talak-KM-2018distributed}
R.~{Talak}, S.~{Karaman}, and E.~{Modiano}, ``Distributed scheduling algorithms
  for optimizing information freshness in wireless networks,'' in \emph{IEEE
  19th International Workshop on Signal Processing Advances in Wireless
  Communications (SPAWC)}, 2018, pp. 1--5.

\bibitem{Talak-KM-wiopt2018perfectCSI}
------, ``Optimizing age of information in wireless networks with perfect
  channel state information,'' in \emph{2018 16th International Symposium on
  Modeling and Optimization in Mobile, Ad Hoc, and Wireless Networks (WiOpt)},
  2018, pp. 1--8.

\bibitem{Talak-KKM-isit2018}
R.~{Talak}, I.~{Kadota}, S.~{Karaman}, and E.~{Modiano}, ``Scheduling policies
  for age minimization in wireless networks with unknown channel state,'' in
  \emph{IEEE International Symposium on Information Theory (ISIT)}, 2018, pp.
  2564--2568.

\bibitem{Maatouk-AE-ITW2018}
A.~{Maatouk}, M.~{Assaad}, and A.~{Ephremides}, ``The age of updates in a
  simple relay network,'' in \emph{2018 IEEE Information Theory Workshop
  (ITW)}, 2018, pp. 1--5.

\bibitem{Yang-AQP-globecom2019}
H.~H. {Yang}, A.~{Arafa}, T.~Q.~S. {Quek}, and H.~V. {Poor}, ``Locally adaptive
  scheduling policy for optimizing information freshness in wireless
  networks,'' in \emph{2019 IEEE Global Communications Conference (GLOBECOM)},
  2019, pp. 1--6.

\bibitem{Buyukates-SU-JCN2019}
B.~{Buyukates}, A.~{Soysal}, and S.~{Ulukus}, ``Age of information in multihop
  multicast networks,'' \emph{Journal of Communications and Networks}, vol.~21,
  no.~3, pp. 256--267, 2019.

\bibitem{Leng-Yener-2019TCCN}
S.~{Leng} and A.~{Yener}, ``Age of information minimization for an energy
  harvesting cognitive radio,'' \emph{IEEE Transactions on Cognitive
  Communications and Networking}, vol.~5, no.~2, pp. 427--439, 2019.

\bibitem{Farazi-KB-JCN2019}
S.~{Farazi}, A.~G. {Klein}, and D.~R. {Brown}, ``Fundamental bounds on the age
  of information in multi-hop global status update networks,'' \emph{Journal of
  Communications and Networks}, vol.~21, no.~3, pp. 268--279, 2019.

\bibitem{Kaul-YG-globecom2011piggybacking}
S.~K. Kaul, R.~D. Yates, and M.~Gruteser, ``On piggybacking in vehicular
  networks,'' in \emph{{IEEE} Global Telecommunications Conference, {GLOBECOM}
  2011}, Dec. 2011.

\bibitem{Bedewy-SS-isit2016}
A.~M. Bedewy, Y.~Sun, and N.~B. Shroff, ``Optimizing data freshness,
  throughput, and delay in multi-server information-update systems,'' in
  \emph{Proc. IEEE Int'l. Symp. Info. Theory (ISIT)}, 2016, pp. 2569--2574.

\bibitem{Bedewy-SS-isit2017}
------, ``Age-optimal information updates in multihop networks,'' in
  \emph{Proc. IEEE Int'l. Symp. Info. Theory (ISIT)}, June 2017, pp. 576--580.

\bibitem{Bedewy-SS-ToN2019}
A.~M. {Bedewy}, Y.~{Sun}, and N.~B. {Shroff}, ``The age of information in
  multihop networks,'' \emph{IEEE/ACM Transactions on Networking}, vol.~27,
  no.~3, pp. 1248--1257, 2019.

\bibitem{Zhong-YS-allerton2017}
J.~Zhong, R.~Yates, and E.~Soljanin, ``Status updates through multicast
  networks,'' in \emph{Proc. Allerton Conf. on Commun., Control and Computing},
  Oct. 2017, pp. 463--469.

\bibitem{Sang-LJ-globecom2017}
Y.~Sang, B.~Li, and B.~Ji, ``The power of waiting for more than one response in
  minimizing the age-of-information,'' in \emph{IEEE Global Communications
  Conference (GLOBECOM)}, Dec 2017.

\bibitem{Zhong-YS-aoi2018}
J.~Zhong, R.~Yates, and E.~Soljanin, ``Minimizing content staleness in
  dynamo-style replicated storage systems,'' in \emph{Infocom Workshop on Age
  of Information}, Apr. 2018, arXiv preprint arXiv:1804.00742.

\bibitem{Zhong-YS-spawc2018}
------, ``Multicast with prioritized delivery: How fresh is your data?'' in
  \emph{{Signal Processing Advance for Wireless Communications (SPAWC)}}, Jun.
  2018, pp. 476--480.

\bibitem{Zhong-Yates-dcc2016lossless}
J.~Zhong and R.~D. Yates, ``Timeliness in lossless block coding,'' in
  \emph{2016 Data Compression Conference (DCC)}, March 2016, pp. 339--348.

\bibitem{Zhong-YS-isit2017}
J.~Zhong, R.~Yates, and E.~Soljanin, ``Backlog-adaptive compression: Age of
  information,'' in \emph{Proc. IEEE Int'l. Symp. Info. Theory (ISIT)}, Jun.
  2017, pp. 566--570.

\bibitem{Mayekar-PT-isit2018lossless}
P.~Mayekar, P.~Parag, and H.~Tyagi, ``Optimal lossless source codes for timely
  updates,'' in \emph{Proc. IEEE Int'l. Symp. Info. Theory (ISIT)}, Jun. 2018,
  pp. 1246--1250.

\bibitem{Mayeker-PT-arxiv2018}
\BIBentryALTinterwordspacing
------, ``Optimal source codes for timely updates,'' \emph{CoRR}, vol.
  abs/1810.05561, 2018. [Online]. Available:
  \url{http://arxiv.org/abs/1810.05561}
\BIBentrySTDinterwordspacing

\bibitem{Costa-VE-icc2015}
M.~Costa, S.~Valentin, and A.~Ephremides, ``On the age of channel information
  for a finite-state {Markov} model,'' in \emph{2015 IEEE International
  Conference on Communications (ICC)}, June 2015, pp. 4101--4106.

\bibitem{Klein-FHB-TW2017}
A.~G. Klein, S.~Farazi, W.~He, and D.~R. Brown, ``Staleness bounds and
  efficient protocols for dissemination of global channel state information,''
  \emph{IEEE Transactions on Wireless Communications}, vol.~16, no.~9, pp.
  5732--5746, Sept 2017.

\bibitem{Farazi-KB-icccn2017}
S.~Farazi, A.~G. Klein, and D.~R. Brown, ``Bounds on the age of information for
  global channel state dissemination in fully-connected networks,'' in
  \emph{Int'l Conference on Computer Communication and Networks (ICCCN)}, July
  2017.

\bibitem{Farazi-KB-icassp2016}
------, ``On the average staleness of global channel state information in
  wireless networks with random transmit node selection,'' in \emph{2016 IEEE
  International Conference on Acoustics, Speech and Signal Processing
  (ICASSP)}, March 2016, pp. 3621--3625.

\bibitem{Bhambay-PP-wcnc2017}
S.~Bhambay, S.~Poojary, and P.~Parag, ``Differential encoding for real-time
  status updates,'' in \emph{2017 IEEE Wireless Communications and Networking
  Conference (WCNC)}, March 2017.

\bibitem{HeFD-aoi2018}
Q.~He, G.~Dan, and V.~Fodor, ``Minimizing age of correlated information for
  wireless camera networks,'' in \emph{IEEE Conference on Computer
  Communications (INFOCOM) Workshops}, April 2018, pp. 547--552.

\bibitem{Abd-Elmagid-Dhillon-vt2019}
M.~A. {Abd-Elmagid} and H.~S. {Dhillon}, ``Average peak age-of-information
  minimization in uav-assisted iot networks,'' \emph{IEEE Transactions on
  Vehicular Technology}, vol.~68, no.~2, pp. 2003--2008, 2019.

\bibitem{Hribar-CKD-globecom2017}
J.~Hribar, M.~Costa, N.~Kaminski, and L.~A. DaSilva, ``Updating strategies in
  the internet of things by taking advantage of correlated sources,'' in
  \emph{IEEE Global Communications Conference (GLOBECOM)}, Dec 2017.

\bibitem{Yates-THR-infocom2017}
R.~Yates, M.~Tavan, Y.~Hu, and D.~Raychaudhuri, ``Timely cloud gaming,'' in
  \emph{Proc. INFOCOM}, May 2017, pp. 1--9.

\bibitem{bastopcu-Ulukus-arxiv2020google}
M.~Bastopcu and S.~Ulukus, ``Who should google scholar update more often?''
  2020.

\bibitem{Nguyen-KKWE-wiopt2017}
G.~D. Nguyen, S.~Kompella, C.~Kam, J.~E. Wieselthier, and A.~Ephremides,
  ``Impact of hostile interference on information freshness: A game approach,''
  in \emph{Int'l Symposium on Modeling and Optimization in Mobile, Ad Hoc, and
  Wireless Networks (WiOpt)}, May 2017.

\bibitem{Xiao-Sun-aoi2018}
Y.~Xiao and Y.~Sun, ``A dynamic jamming game for real-time status updates,'' in
  \emph{IEEE Conference on Computer Communications (INFOCOM) Workshops}, April
  2018, pp. 354--360.

\bibitem{Gopal-Kaul-aoi2018}
S.~Gopal and S.~K. Kaul, ``A game theoretic approach to {DSRC} and {WiFi}
  coexistence,'' in \emph{IEEE Conference on Computer Communications (INFOCOM)
  Workshops}, April 2018, pp. 565--570.

\bibitem{Garnaev-ZZY-aoi2019}
A.~Garnaev, J.~Zhong, W.~Zhang, and R.~Yates, ``Maintaining information
  freshness under jamming,'' in \emph{Infocom Workshop on Age of Information},
  Apr. 2019.

\bibitem{Hespanha-2006modelling}
J.~Hespanha, ``Modelling and analysis of stochastic hybrid systems,'' \emph{IEE
  Proceedings-Control Theory and Applications}, vol. 153, no.~5, pp. 520--535,
  2006.

\bibitem{Vermes-1980}
D.~Vermes, ``Optimal dynamic control of a useful class of randomly jumping
  processes,'' International Institute for Applied Systems Analysis, Tech. Rep.
  PP-80-015, 1980.

\bibitem{Gnedenko-Kovalenko-1966}
B.~Gnedenko and I.~Kovalenko, ``Introduction to the theory of mass service,''
  \emph{Nau (trad, ingles: 1968, Jerusalem)}, 1966.

\bibitem{Davis-1984}
M.~H.~A. Davis, ``Piecewise-deterministic {Markov} processes: a general class
  of nondiffusion stochastic models,'' \emph{J. Roy. Statist. Soc.}, vol.~46,
  pp. 353--388, 1984.

\bibitem{Deville-DDZ-siam2016moment}
L.~DeVille, S.~Dhople, A.~D. Dom{\'\i}nguez-Garc{\'\i}a, and J.~Zhang, ``Moment
  closure and finite-time blowup for piecewise deterministic {Markov}
  processes,'' \emph{SIAM Journal on Applied Dynamical Systems}, vol.~15,
  no.~1, pp. 526--556, 2016.

\bibitem{Yates-aoi2018}
R.~D. Yates, ``Age of information in a network of preemptive servers,'' in
  \emph{IEEE Conference on Computer Communications (INFOCOM) Workshops}, Apr.
  2018, pp. 118--123, arXiv preprint arXiv:1803.07993.

\bibitem{Teel-SS-2014stability}
A.~R. Teel, A.~Subbaraman, and A.~Sferlazza, ``Stability analysis for
  stochastic hybrid systems: A survey,'' \emph{Automatica}, vol.~50, no.~10,
  pp. 2435--2456, 2014.

\bibitem{Hespanha-course}
J.~P. Hespanha, ``{Hybrid and Switched Systems: ECE 229 -- Fall 2005},''
  \url{https://www.ece.ucsb.edu/~hespanha/ece229/}.

\bibitem{Seneta-1981non-negative}
E.~Seneta, \emph{Non-negative Matrices and Markov Chains}, 2nd~ed.\hskip 1em
  plus 0.5em minus 0.4em\relax Springer Verlag, 1981.

\end{thebibliography}
\end{spacing}

\begin{IEEEbiographynophoto}{Roy D. Yates} is a Distinguished Professor in the ECE Department and the Wireless Information Network Laboratory at Rutgers University. He is a co-author of
the textbook {\em Probability and Stochastic Processes: A Friendly Introduction for Electrical and Computer Engineers} published by John Wiley and Sons. He received the B.S.E. degree in 1983 from Princeton and the S.M. and Ph.D. degrees in 1986 and 1990 from MIT, all in Electrical Engineering. He was an associate editor of the IEEE TRANSACTIONS ON INFORMATION THEORY in 2009-12.   He received the 2003 IEEE Marconi Prize Paper Award in Wireless Communications, a best paper award at the ICC 2006 Wireless Communications Symposium, and the 2011 Rutgers University Teacher-Scholar award. 
\end{IEEEbiographynophoto}

\end{document}